\documentclass[runningheads]{llncs}
\usepackage[T1]{fontenc}
\usepackage{graphicx}
\usepackage{amsmath}
\usepackage{amssymb}
\usepackage{subfig}
\usepackage{float}
\usepackage{booktabs}
\usepackage{subcaption}
\usepackage{xspace}
\usepackage{dsfont}
\usepackage{xcolor}
\usepackage{tikz}
\usetikzlibrary{automata, positioning, arrows.meta}
\usepackage{longtable}
\usepackage{hyperref} 
\usepackage{cleveref}

\begin{document}
\title{Scenario-Based Compositional Statistical Model Checking for Safety Specifications}
\titlerunning{Scenario-Based Compositional SMC for Safety Specifications}
%
\author{
Abhinav Pomalapally
\and
Arya Raeesi
\and
\\
Kevin Kai-Chun Chang
\and
Beyazit Yalcinkaya
\and
Sanjit A. Seshia
}
\authorrunning{A. Pomalapally and A. Raeesi et al.}
%
\institute{University of California, Berkeley, USA
\email{\{abhinav\_pomalapally,aryaraeesi\}@berkeley.edu}
\\
\email{\{kaichunchang,beyazit,sseshia\}@berkeley.edu}
}

%
\maketitle              
\begin{abstract}
In safety-critical domains such as autonomous driving, systems must be evaluated across a large number of environment conditions, often represented as \emph{composite scenarios} built from \emph{primitive scenarios}. Existing statistical model checking (SMC) approaches analyze each composite scenario independently, requiring many expensive simulations and resulting in substantial redundant computation when scenarios share common structure. This work introduces a scenario-based compositional SMC framework for safety and co-safety specifications, enabling efficient analysis of composite scenarios. Our approach decomposes scenarios into primitives and specifications into sub-specifications, verifies each primitive independently, and composes the resulting statistical estimates using importance sampling and kernel density estimation.
Our empirical evaluation shows that the proposed framework can accurately answer verification queries for previously unseen composite scenarios while reducing simulation cost through parallelization and trace~reuse.
\end{abstract}

\newcommand{\straight}{\texttt{S}\xspace}
\newcommand{\intersection}{\texttt{X}\xspace}
\newcommand{\roundabout}{\texttt{O}\xspace}
\newcommand{\curved}{\texttt{C}\xspace}

\newcommand{\seq}{\text{$\rightarrow$}\xspace}
\newcommand{\cho}{\texttt{choose}\xspace}
\newcommand{\shu}{\texttt{shuffle}\xspace}

\newcommand{\ind}{\mathds{1}\xspace}

\newcommand{\red}[1]{\textcolor{red}{#1}}
\newcommand{\blue}[1]{\textcolor{blue}{#1}}
\newcommand{\green}[1]{\textcolor{green}{#1}}


\section{Introduction}\label{sec:intro}

Artificial intelligence (AI) systems are increasingly deployed in safety-critical domains, including autonomous driving, robotics, healthcare, and cyber-physical systems in general.
As these systems become more pervasive, ensuring their safety and reliability remains a fundamental challenge. However, the complexity of modern AI-based systems renders traditional verification techniques largely inapplicable.
Classical formal verification methods typically require an explicit and tractable model of the system under analysis, an assumption that is often violated in contemporary AI systems due to their inherent opacity.

A practical alternative is to treat the system under analysis as a black box and reason about its behavior through simulation-based analysis~\cite{seshia2022toward}.
This approach is particularly prevalent in autonomous driving, where validating a system through exhaustive real-world testing is infeasible.
Instead, developers evaluate autonomous driving systems using a corpus of \emph{scenarios}, either manually designed by domain experts or extracted from real-world driving data, and assess the system's performance across these scenarios through repeated simulations.

Statistical model checking (SMC) provides a principled framework for such simulation-based verification~\cite{legay2010statistical}.
SMC is a formal verification technique that determines whether a stochastic system satisfies a given property within a particular scenario by analyzing a finite set of execution traces using statistical methods such as hypothesis testing or confidence interval estimation. By mitigating the state-space explosion associated with exhaustive model checking, SMC enables verification of systems whose complexity would otherwise make formal analysis intractable. Nevertheless, SMC continues to face significant scalability challenges. Obtaining meaningful and high-confidence verification results often requires executing a large number of computationally expensive, long-running simulations. Moreover, SMC must typically be repeated whenever the scenario defining the environment conditions changes, leading to redundant computation.

In this paper, we introduce a scenario-based compositional SMC framework for safety and co-safety specifications, i.e., specifications whose violation or satisfaction can be determined from finite executions. Our approach leverages both scenario and specification decomposition. Given a \emph{composite scenario}, we first decompose it into a collection of \emph{primitive scenarios}. Simultaneously, we decompose the target safety specification into a set of sub-specifications. The black-box system is then evaluated independently on each primitive scenario with respect to all relevant sub-specifications. These verification tasks can be executed in parallel and require shorter simulation runs than the original monolithic verification problem. Finally, we compose the resulting statistical estimates to obtain a verification result for the original composite scenario and specification.
Furthermore, our compositional framework enables the reuse of traces generated for primitive scenarios across multiple composite scenarios, thereby avoiding redundant simulation effort and significantly reducing overall verification cost.

We summarize our contributions as follows:
(i) we introduce a novel compositional statistical model checking framework for safety and co-safety specifications;
(ii) we implement the proposed framework and integrate it into VerifAI~\cite{dreossi2019verifai}, a toolkit for the formal design and analysis of AI-based systems; and
(iii) we conduct an empirical evaluation of the proposed approach and demonstrate its effectiveness in improving verification efficiency while maintaining accuracy.

\subsubsection{Related Work.} 
To address the verification of AI-based systems, simulation-based formal analysis has been widely adopted~\cite{seshia2022toward,dreossi2019verifai,abbas2013probabilistic}. Specifically, statistical model checking (SMC)~\cite{sen2004statistical,sen2005statistical,younes2006numerical,younes2006statistical,legay2010statistical,clarke2011statistical,zuliani2013bayesian,agha2018survey} mitigates the state-space explosion of exact formal methods by evaluating a finite set of execution traces, providing probabilistic guarantees bounded by statistical confidence intervals. Seshia et al.~\cite{seshia2022toward} advocate for a compositional approach to scale up verification of AI systems. Classic compositional verification (e.g.,~\cite{clarke1989compositional,cobleigh2003learning,chilton2014compositional,junges2022abstraction,watanabe2023compositional}) is not directly applicable to simulation-based verification of AI systems. Existing compositional approaches for AI systems vary in their axes of decomposition; while some frameworks decompose internal system architectures or neural network layers~\cite{vin2025scenicprover,duong2026compositional}, our work focuses on the spatial and temporal decomposition of environmental scenarios and specifications.

Dreossi et al.~\cite{dreossi2019compositional} has proposed a framework combining compositional analysis with simulation-based verification, but it focuses strictly on component-level decomposition and targets active falsification rather than SMC.
The closest prior work to the proposed framework introduces a compositional approach to SMC~\cite{yalcinkaya2023compositional}.
However, it is strictly restricted to Markovian, i.e., memoryless, specifications. Our new framework extends compositional SMC to non-Markovian safety and co-safety specifications, overcoming a fundamental limitation.

Another related line of work explores scenario decomposition, where environments and specifications are divided into primitive segments either spatially~\cite{watson2025scenario,ivanov2021compositional} or temporally~\cite{chang2024dynamic}. However, these differ significantly in their verification methods: they rely on explicit white-box access~\cite{ivanov2021compositional}, offline probabilistic abstractions~\cite{watson2025scenario}, or remain confined to falsification~\cite{chang2024dynamic}. In contrast, our framework achieves black-box compositional SMC for safety and co-safety properties.

\section{Motivating Example}\label{sec:example}

   \label{fig:example} 

\par\medskip
\begingroup
\captionsetup{type=figure}
\captionsetup[subfloat]{justification=centering}
\centering
\subfloat[\straight\label{fig:s}]{%
       \includegraphics[width=0.24\linewidth,height=75pt,keepaspectratio]{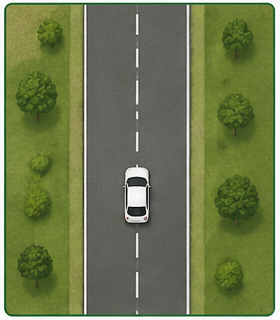}}\hfill
\subfloat[\intersection\label{fig:x}]{%
       \includegraphics[width=0.24\linewidth,height=75pt,keepaspectratio]{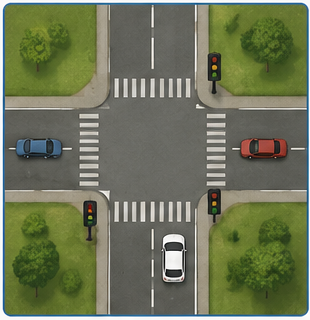}}\hfill
\subfloat[\roundabout\label{fig:o}]{%
       \includegraphics[width=0.24\linewidth,height=75pt,keepaspectratio]{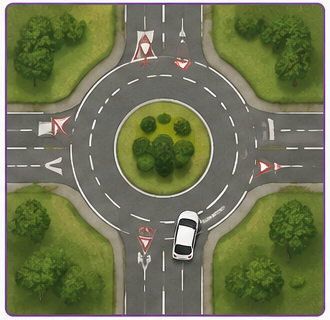}}\hfill
\subfloat[\curved\label{fig:c}]{%
       \includegraphics[width=0.24\linewidth,height=75pt,keepaspectratio]{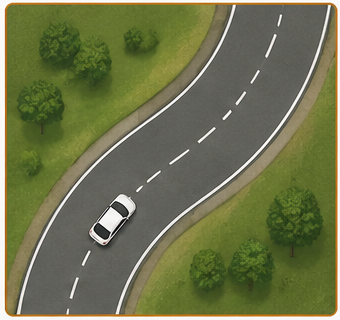}}
\captionof{figure}{Primitive scenarios considered throughout the paper: a
straight road segment (\straight), an intersection (\intersection), a
roundabout (\roundabout), and a curved road segment (\curved).}
\label{fig:example}
\endgroup
\par\medskip


Consider evaluating an autonomous vehicle across various driving scenarios (\Cref{fig:example}). A common approach to constructing such scenarios is to use a set of simpler primitive scenarios as building blocks for defining more complex composite scenarios. Suppose that our primitive scenarios consist of a straight road segment (\straight), an intersection (\intersection), a roundabout (\roundabout), and a curved road segment (\curved). Scenario description languages such as Scenic~\cite{fremont2019scenic,fremont2023scenic} support this style of programmatic scenario construction through several composition operators. \emph{Sequential composition} specifies the exact order in which scenarios are encountered (e.g., \straight \seq \intersection). \emph{Choice composition} randomly selects one scenario from a set of alternatives (e.g., $\cho(\straight, \intersection, \roundabout, \curved)$). Finally, \emph{shuffle composition} randomly permutes the order of a collection of scenarios (e.g., $\shu(\straight, \intersection, \roundabout, \curved)$).

Our goal is to efficiently test the autonomous vehicle in composite scenarios against safety and co-safety specifications, i.e., temporal specifications whose satisfaction or violation can be determined from a finite execution.
%
A monolithic application of SMC treats each composite scenario independently. For every new scenario and specification pair, SMC executes a large number of simulations and analyzes the resulting traces. As the complexity of composite scenarios grows, this process quickly becomes computationally expensive, requiring repeated execution of long-running simulations that often share substantial~structure.

Our approach avoids this redundancy by shifting the computational effort to the primitive scenarios. Instead of directly evaluating every composite scenario, we execute the system independently within each primitive scenario and store the resulting traces. We then evaluate these traces against all sub-specifications induced by the target specification, corresponding to the sub-automata of its monitor automaton. Given a composite scenario, we compose the previously computed results to estimate the probability of satisfying the target specification.

A key challenge is that the distribution of behaviors observed in an isolated primitive scenario may differ from the distribution induced when that same primitive appears as part of a larger composite scenario. To account for this distributional mismatch, we employ importance sampling~\cite{kloek1978bayesian} together with kernel density estimation (KDE). Intuitively, traces generated from primitive scenarios are weighted according to their likelihood under the target composite scenario. This enables us to estimate results for previously unseen composite scenarios without actually running the system in those scenarios. Consequently, our approach generates computationally efficient and highly accurate estimates for SMC queries.

\section{Method}\label{sec:method}

In this section, we present our compositional approach for checking safety specifications.
We assume that the controller under consideration is Markovian, i.e., it outputs actions that solely depend on the current state of the plant, and that it operates in a Markov Decision Process (MDP).
A \emph{primitive scenario} defines environment conditions from a set of initial states to a set of terminal states, e.g., the primitive scenario in \Cref{fig:x} starts at the car's initial position and ends when the car leaves the intersection.
A \emph{composite scenario} is a composition of such primitives, e.g., \straight \seq \intersection sequentially composes \straight and \intersection, defining a scenario that starts at the beginning of a straight road and ends after an intersection.

Let $\tau \sim (\pi, \mathcal{S})$ denote a (labeled) trace sampled by running a policy $\pi$ in a scenario $\mathcal{S}$, let $\mathcal{T}$ be a finite set of such traces, and let $\phi$ be a safety specification in automaton form.
SMC estimates the specification satisfaction probability~as:
\begin{align}\label{eqn:smc}
\Pr[\tau \models \phi]
=
\frac{1}{|\mathcal{T}|}
\sum_{\tau_i \in \mathcal{T}} \ind\{\tau_i \models \phi\},
\end{align}
where $\tau \models \phi$ denotes that the trace $\tau$ satisfies the safety specification $\phi$.
Note that $|\mathcal{T}|$ can be chosen via Hoeffding's inequality to achieve a desired confidence~level.

We rewrite this expression in terms of the \emph{state distribution} of the safety specification.
Observe that for a safety specification $\phi = (Q, \Sigma, \delta, q_0, F)$, its state distribution $\mu_0 \in \Delta(Q)$ is initially concentrated at $q_0$, i.e., $\mu_0(q_0) = 1$, as $q_0$ is the initial state.
Let $\mu_\mathcal{S}$ denote the state distribution induced by transitioning the automaton using traces $\tau \sim (\pi, \mathcal{S})$.
$\mu_\mathcal{S}$ summarizes execution history within each trace relevant to the specification, allowing our method to handle non-Markovian temporal properties through Markovian updates over automaton states.
Therefore, we can rewrite \Cref{eqn:smc} as follows:
\begin{align}\label{eqn:mu}
\Pr[\tau \models \phi]
=
\sum_{q \in Q} \ind\{\delta^*(q, \tau) \in F\} \mu_0(q)
=
\sum_{q \in Q} \ind\{q \in F\} \mu_\mathcal{S}(q),
\end{align}
where $\delta^*$ is the transitive closure of the automaton transition function $\delta$.
Observe that the expression in the middle marginalizes over all initial states using the initial state distribution $\mu_0$, and the expression on the right marginalizes over all final states using the resulting output state distribution $\mu_\mathcal{S}$.
Let $\mathcal{T}_{q \to q'}$ denote the subset of the given traces $\mathcal{T}$ that ends at an automaton state $q'$ starting from an automaton state $q$.
We can estimate $\mu_\mathcal{S}$ using $\mu_0$ and $\mathcal{T}$ as follows:
\begin{align}\label{eqn:est}
    \mu_\mathcal{S}(q')
    =
    \sum_{q \in Q}
    \mu_0(q)
    \frac{1}{|\mathcal{T}_{q \to q'}|}
    \sum_{\tau_i \in \mathcal{T}_{q \to q'}}
    \ind\{\delta^*(q, \tau_i) = q'\}.
\end{align}
This expression shows the one-step propagation rule for computing the resulting automaton state distribution after a scenario $\mathcal{S}$. We emphasize that \Cref{eqn:est} is not intended as a novel derivation; rather, it provides the empirical estimation needed to obtain the automaton state distribution induced by a scenario from the initial distribution and available traces.
We use this idea to compositionally compute SMC queries for sequential, choose, and shuffle compositions.

As co-safety specifications are the complement of safety specifications, given a co-safety specification $\phi$, we take its complement $\bar \phi$ which is a safety specification, run SMC, and take the complement of the resulting satisfaction probability, i.e.,
\begin{align}
    \Pr[\tau \models \phi] = 1 - \Pr[\tau \models \bar\phi],
\end{align}
where $\bar\phi$ is a safety specification, and $\Pr[\tau \models \bar\phi]$ is computed as in \Cref{eqn:smc}.



\subsection{Sequential Composition}\label{sec:seq}

Consider the sequential composition \straight \seq \intersection.
Let $\mathcal{T}^{\straight}$ and $\mathcal{T}^{\intersection}$ denote traces independently generated for the primitive scenarios \straight and \intersection, respectively.
Our goal is to use the primitive scenario traces $\mathcal{T}^{\straight}$ and $\mathcal{T}^{\intersection}$ to estimate the satisfaction probability for the composite scenario \straight \seq \intersection without actually generating traces for it.
Let $\tau^{\straight \seq \intersection}$, $\tau^{\straight}$, and $\tau^{\intersection}$ denote traces sampled for \straight \seq \intersection, \straight, and \intersection, respectively.
Let $\mu_0$ be the initial automaton state distributions, and let $\mu_{\straight \seq \intersection}$, $\mu_{\straight}$, and $\mu_{\intersection}$ denote the automaton state distribution induced by the scenarios \straight \seq \intersection, \straight, and \intersection, respectively.
Using \Cref{eqn:mu}, we can compute the SMC query~as~follows:
\begin{align}
\Pr[\tau^{\straight \seq \intersection} \models \phi]
=
\sum_{q \in Q} \ind \{ q \in F \} \mu_{\straight \seq \intersection} (q).
\end{align}
We can estimate $\mu_{\straight \seq \intersection}$ using $\mu_{\straight}$ and \Cref{eqn:est} as follows:
\begin{align}\label{eqn:seq}
    \mu_\mathcal{\straight \seq \intersection}(q')
    =
    \sum_{q \in Q}
    \mu_{\straight}(q)
    \frac{1}{|\mathcal{T}^{\intersection}_{q \to q'}|}
    \sum_{\tau_i \in \mathcal{T}^{\intersection}_{q \to q'}}
    \ind\{\delta^*(q, \tau_i) = q'\}
    \frac{o_{\straight,q}(\tau_i)}{\iota_{\intersection}(\tau_i)},
\end{align}
where $\mu_{\straight}$ is estimated recursively using $\mu_0$ and \Cref{eqn:est}.
Here, notice that $\mathcal{T}^{\intersection}$ is generated using the initial MDP state distribution of \intersection, denoted by $\iota_\intersection$, and the result we want is with respect to the induced output MDP state distribution of \straight for traces that end at the automaton state $q$ (since we are marginalizing over $q$), denoted by $o_{\straight, q}$.
Therefore, we perform \emph{importance sampling} to account for this distributional mismatch, given by the fraction at the end of the expression.
This allows us to treat traces in $\mathcal{T}^{\intersection}$ as if they were generated starting from the resulting output MDP state distribution of the preceding scenario $\straight$.
Using this approach, we start from the initial state distribution and compute its evolution throughout the given sequential composition, i.e., $\mu_0 \xrightarrow{\straight} \mu_{\straight} \xrightarrow{\intersection} \mu_{\straight \seq \intersection}$.

\Cref{eqn:seq} is a valid decomposition of the monolithic SMC query due to the assumed Markovian dynamics of the underlying stochastic~process, i.e., it is sufficient to account for the distributional mismatch between the last and the first MDP states of the scenarios \straight and \intersection, respectively.
In practice, it is crucial to ensure that the output and input MDP state distributions of sequentially composed scenarios like \straight and \intersection overlap.
We assume it to be the case, and thus, we only account for the distributional mismatch through importance sampling.
As such, we estimate $\iota_{\intersection}$ and $o_{\straight}$ using Gaussian \emph{kernel density estimation} (KDE), an asymptotically valid nonparametric method.
KDE performs well in moderate dimensions, whereas it might suffer from the curse of dimensionality in high-dimensional spaces.
We expect to apply the proposed approach in moderately sized semantic feature spaces; therefore, for our purposes, KDE suffices.

\subsection{Choose Composition}

Consider the choose composition $\cho(\straight, \intersection)$.
Let $\tau^{\cho(\straight, \intersection)}$, $\tau^{\straight}$, and $\tau^{\intersection}$ denote traces sampled for the composite scenario and the corresponding primitive scenarios.
The satisfaction probability can be written as a mixture distribution:
\begin{align}\label{eqn:choose}
\Pr[\tau^{\cho(\straight, \intersection)} \models \phi]
=
\frac{1}{2}
\Pr[\tau^{\straight} \models \phi]
+
\frac{1}{2}
\Pr[\tau^{\intersection} \models \phi].
\end{align}
Intuitively, the composite scenario induces a probabilistic choice over the constituent scenarios, and the overall satisfaction probability is obtained by marginalizing over the selected branch.
This expression can be easily extended to $n$-ary and weighted choose compositions by marginalizing with branch probabilities.



\subsection{Shuffle Composition}

Finally, consider the shuffle composition $\shu(\straight, \intersection)$, which randomly picks a permutation of the given scenarios to sequentially compose.
Let $\tau^{\shu(\straight, \intersection)}$ denote a trace sampled from it.
We can decompose the SMC query as follows:
\begin{align}\label{eqn:shuffle}
\Pr[\tau^{\shu(\straight, \intersection)} \models \phi]
=
\frac{1}{2}
\Pr[\tau^{\straight \seq \intersection} \models \phi]
+
\frac{1}{2}
\Pr[\tau^{\intersection \seq \straight} \models \phi].
\end{align}
This can be trivially extended to $n$-ary and weighted shuffle compositions.
While shuffle composition results in a factorial growth in the number of permutations, we expect the number of shuffled scenarios to remain relatively small.

\section{Empirical Evaluation}\label{sec:exps}


\tikzset{
  specdfa/.style={
    shorten >=1pt, node distance=1.2cm and 1.35cm, on grid, auto,
    >={Stealth[length=4pt]},
    every state/.style={font=\scriptsize, inner sep=1pt, minimum size=14pt},
    every edge/.style={draw, ->},
    font=\tiny,
    initial text={},
  },
}

In this section, we empirically evaluate the proposed framework along two dimensions: \emph{accuracy} and \emph{efficiency}. To evaluate accuracy, we impose a fixed \textit{trace budget} on both our compositional SMC approach and a monolithic SMC baseline, and compare the resulting satisfaction-probability estimates against those obtained by the baseline.
To evaluate efficiency, we impose a fixed \textit{time budget}, limiting both approaches to generating traces within the allotted time. We then compare the resulting estimates with their respective counterparts from the fixed-trace-budget experiment, assessing whether each approach can converge to accurate estimates under a limited data budget. This setting is particularly relevant during early-stage system development, when rapid feedback is valuable.

We implemented the proposed framework\footnote{Our code is at \href{https://github.com/abhip02/VerifAI/tree/compositional-analysis}{https://github.com/abhip02/VerifAI/tree/compositional-analysis}.} as a prototype integrated with VerifAI~\cite{dreossi2019verifai}, a toolkit for the formal design and analysis of AI-based systems.
We use MetaDrive~\cite{li2022metadrive} as the simulator and Scenic~\cite{fremont2019scenic,fremont2023scenic} as the scenario description language.
We test MetaDrive’s built-in expert policy, a neural controller trained to handle general driving scenarios.

We evaluate the framework on seven composite scenarios:
(i) \straight\seq\intersection,
(ii) \straight\seq\intersection\seq\straight,
(iii) \straight\seq\roundabout\seq\curved,
(iv) \curved\seq\straight\seq\intersection\seq\straight,
(v) \curved\seq\intersection\seq\straight\seq\intersection\seq\curved,
(vi) \straight \seq \cho(\curved, \intersection, \roundabout), and 
(vii) \straight \seq \shu(\curved, \intersection, \roundabout).
These comprise five sequential compositions of varying lengths and two compositions using \cho and \shu. We consider four specifications: two safety properties, \textbf{2-Stop}, which forbids more than one near-stop event, and \textbf{Tollgate}, which forbids remaining below a slow-speed threshold for more than three consecutive steps; and two co-safety properties, \textbf{V-Shaped}, which requires the ego to eventually complete a fast--slow--fast speed profile, and \textbf{Sustained}, which requires the ego to eventually sustain a large steering input for three consecutive steps. The four specifications are detailed in~\Cref{appendix:formal_specifications}. This yields $28$ scenario--specification pairs, each evaluated in both experiments.



\subsection{Evaluating Accuracy}
\label{subsec:exp-accuracy} 

In the fixed-trace-budget experiment, we generate $1000$ composite scenario traces for the monolithic SMC baseline. For our compositional SMC method, we instead generate $1000$ traces for each primitive scenario and reuse these traces across all considered composite scenarios.

\begin{table}[t]
\centering
\setlength{\tabcolsep}{10pt}
\caption{Estimated satisfaction probabilities for the \textbf{Tollgate} specification under the fixed-trace-budget regime.}
\label{tab:tollgate-fixed-trace-budget}
\begin{tabular}{l ccc}
\toprule
Scenario & \shortstack{Monolithic\\SMC (baseline)} & \shortstack{Compositional\\SMC (ours)} & $\Delta$
\\
\midrule
\straight\seq\intersection                                          & $0.6750$ & $0.7074$ & $+0.0324$ \\
\straight\seq\intersection\seq\straight                             & $0.6520$ & $0.6898$ & $+0.0378$ \\
\straight\seq\roundabout\seq\curved                                 & $0.4310$ & $0.4740$ & $+0.0430$ \\
\curved\seq\straight\seq\intersection\seq\straight                  & $0.5670$ & $0.6206$ & $+0.0536$ \\
\curved\seq\intersection\seq\straight\seq\intersection\seq\curved   & $0.3820$ & $0.3819$ & $-0.0001$ \\
\straight \seq \cho(\curved, \intersection, \roundabout)            & $0.9800$ & $0.9584$ & $-0.0216$ \\
\straight \seq \shu(\curved, \intersection, \roundabout)            & $0.9150$ & $0.8941$ & $-0.0209$ \\
\bottomrule
\end{tabular}
\end{table}



Table~\ref{tab:tollgate-fixed-trace-budget} presents results for the \textbf{Tollgate} specification as an example. The compositional estimates closely track the values reported by the monolithic baseline, demonstrating the accuracy of our compositional framework.
We observe similar agreement for nearly all remaining scenario--specification pairs, as reported in \Cref{tab:fixed-trace-budget-full} of \Cref{appendix:fixed_budget_results}. One notable exception is the \textbf{Sustained} specification for the \straight \seq \shu(\curved, \intersection, \roundabout) scenario, for which $\Delta$ reaches $+0.1162$.
This gap arises from MDP initial-state correlation discussed in \Cref{sec:seq}: in the monolithic execution, each subsequent scenario begins from an MDP state the previous scenario left off at, whereas the compositional model samples each subsequent scenario from its own initial MDP state distribution.
Although we use importance sampling to account for this distributional mismatch, the quality of the compositional estimate depends directly on the degree of overlap between the MDP state distributions at the interfaces of consecutive scenarios.
This highlights that the initial state distributions of primitive scenarios must be chosen carefully to ensure accurate compositional estimates.
One way we ensure this in our experiments is by simulating distinct geometries of the same primitive scenario to yield a representative state distribution.
For example, we randomly sample four-way intersections on the same map with varying initial velocities and headings.
By diversifying these initial conditions, we maximize the likelihood that state distributions overlap across scenario boundaries.
Finally, note that the expert policy used in our experiments was not trained to satisfy any of the considered specifications; rather, it serves as a placeholder.

\subsection{Evaluating Efficiency}
\label{subsec:eval_eff}

\begin{figure}[t]
    \centering
    \includegraphics[width=0.575\linewidth]{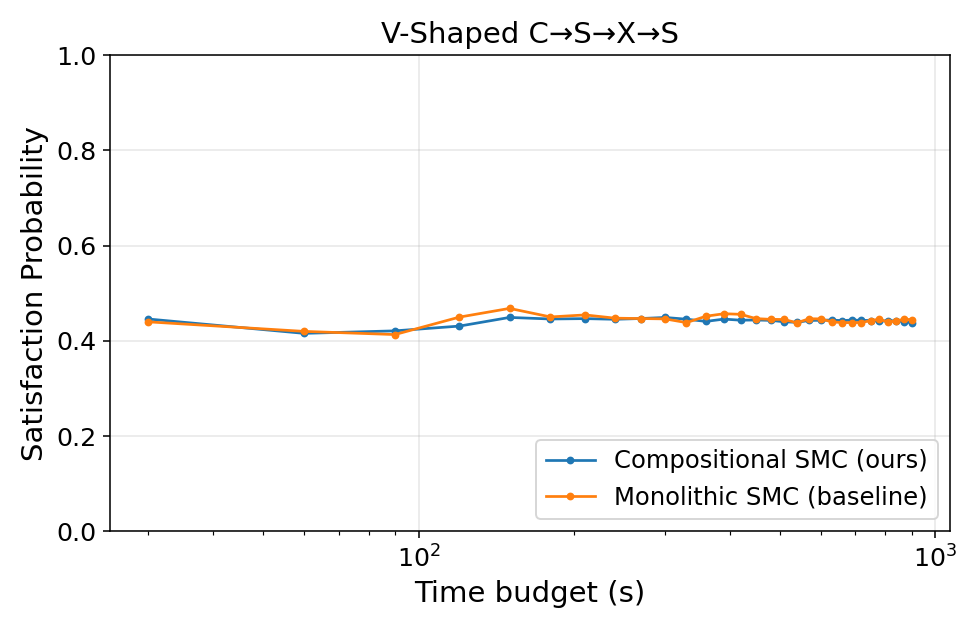}
    \caption{Fixed-time-budget convergence for the \textbf{V-Shaped} specification on the
    \curved\seq\straight\seq\intersection\seq\straight scenario. $753$ monolithic and $14871$ compositional traces generated within the 15-minute budget.}
    \label{fig:time_vshape_CSXS}
    \vspace{-6mm}
\end{figure}

In the fixed-time-budget experiment, both the compositional approach and the monolithic baseline are given a 15-minute wall-clock budget for trace generation. Under this budget, the compositional estimates converge faster toward their respective fixed-trace-budget references than the monolithic baseline, with a mean deviation of $0.029$ versus $0.039$. This improvement is enabled by the compositional method generating $14.6\times$ more traces on average (ranging from $5.2\times$ to $32.3\times$ more traces). \Cref{fig:time_vshape_CSXS} shows a representative scenario--specification pair. Within the same wall-clock budget, the compositional method generates $14{,}871$ primitive traces, compared with $753$ monolithic traces. Full results are provided in  \Cref{app:time-budget-15}.

The gap is concentrated among pairs involving \cho/\shu scenarios, where the per-trace cost is sufficiently high that the 15-minute budget yields as few as $20$ monolithic traces. The gap is largest for the $\langle$\textbf{V-Shaped}, \straight \seq \shu(\curved, \intersection, \roundabout)$\rangle$ pair, for which the monolithic estimate has an error of $0.275$ relative to the reference, compared with $0.014$ for the compositional approach. Even when the time budget is extended to $60$ minutes, the monolithic baseline still achieves an error of $0.188$, substantially higher than the error achieved by the compositional approach with only one quarter of the time budget ($15$ minutes). Full results for the $60$-minute experiment are provided in \Cref{app:time-budget-60}.

Finally, we conduct a compute-matched ablation in which the monolithic baseline retains the 15-minute budget, while the compositional approach receives an equal budget for each of the four primitive scenarios (\straight, \intersection, \roundabout, \curved), yielding $225$ seconds per primitive. With the sample-count advantage removed, the two approaches are essentially tied, with mean deviations of $0.038$ and $0.039$, respectively, across the $28$ scenario--specification pairs. This result further validates that the primary benefit of our approach comes from sample reuse: a single pool of primitive traces can serve all composite scenarios, whereas the monolithic baseline must regenerate traces independently for each composite scenario. Full results are provided in \Cref{app:ablation-matched}.

\section{Conclusion}
We presented a scenario-based compositional statistical model checking framework for safety and co-safety specifications.
By decomposing composite scenarios into primitive scenarios and specifications into sub-specifications, our approach enables independent verification of primitive scenarios whose results are then composed using importance sampling and kernel density estimation to answer SMC queries about previously unseen composite scenarios.
The framework supports sequential, choose, and shuffle compositions.
Our empirical evaluation demonstrates that the proposed approach produces accurate estimates while reducing simulation cost through parallelization and trace reuse---a particularly valuable property in early-stage development when rapid feedback is~essential.

\begin{credits}
\subsubsection{\ackname} 
This work was partially supported by the DARPA TIAMAT program, NSF grant 2303564, the California Partners for Advanced Transportation Technology (PATH) Future Mobility Research Center (FMRC), and Nissan through the iCyPhy Center. This work originated as a class project in EECS 219C at UC Berkeley. 
\end{credits}

%
%

%
\bibliographystyle{splncs04}
\bibliography{references}

\appendix
\crefalias{section}{appendix}
\crefalias{subsection}{appendix}
\crefname{appendix}{Appendix}{Appendices}
\Crefname{appendix}{Appendix}{Appendices}
\section{Formal Specifications}\label{appendix:formal_specifications}

\raggedbottom

\begin{figure}[!htbp]
    \centering
    \captionsetup[subfloat]{justification=centering}
  \subfloat[\textbf{2-Stop} (safety)\label{fig:spec_two_stop}]{%
    \begin{tikzpicture}[specdfa]
      \node[state, initial, accepting] (q0) {$q_0$};
      \node[state, accepting, right=of q0] (q1) {$q_1$};
      \node[state, right=of q1] (qb) {$q_\bot$};
      \path (q0) edge[loop above] node {$v \ge 3.5$} (q0)
            (q0) edge node {$v < 3.5$} (q1)
            (q1) edge[loop above] node {$v \ge 3.5$} (q1)
            (q1) edge node {$v < 3.5$} (qb)
            (qb) edge[loop above] node {$\ast$} (qb);
    \end{tikzpicture}}
    \hfill%
  \subfloat[\textbf{Tollgate} (safety)\label{fig:spec_tollgate}]{%
    \begin{tikzpicture}[specdfa]
      \node[state, initial, accepting] (q0) {$q_0$};
      \node[state, accepting, right=of q0] (s1) {$q_1$};
      \node[state, accepting, right=of s1] (s2) {$q_2$};
      \node[state, right=of s2] (qb) {$q_\bot$};
      \path (q0) edge[loop above] node {$v \ge 3.5$} (q0)
            (q0) edge node {$v < 3.5$} (s1)
            (s1) edge node {$v < 3.5$} (s2)
            (s2) edge node {$v < 3.5$} (qb)
            (s1) edge[bend left=20]node[below] {$v \ge 3.5$}  (q0)
            (s2) edge[bend left=70] node[below] {$v \ge 3.5$} (q0)
            (qb) edge[loop above] node {$\ast$} (qb);
    \end{tikzpicture}}
\\
  \subfloat[\textbf{V-Shaped} (co-safety)\label{fig:spec_v_shaped}]{%
    \begin{tikzpicture}[specdfa]
      \node[state, initial] (q0) {$q_0$};
      \node[state, right=of q0] (q1) {$q_1$};
      \node[state, right=of q1] (q2) {$q_2$};
      \node[state, accepting, right=of q2] (qf) {$q_F$};
      \path (q0) edge[loop above] node {$v < 7$} (q0)
            (q0) edge node {$v \ge 7$} (q1)
            (q1) edge[loop above] node {$v > 3.5$} (q1)
            (q1) edge node {$v \le 3.5$} (q2)
            (q2) edge[loop above] node {$v < 7$} (q2)
            (q2) edge node {$v \ge 7$} (qf)
            (qf) edge[loop above] node {$\ast$} (qf);
    \end{tikzpicture}}
    \hfill%
  \subfloat[\textbf{Sustained} steering (co-safety)\label{fig:spec_sustained}]{%
    \begin{tikzpicture}[specdfa]
      \node[state, initial] (q0) {$q_0$};
      \node[state, right=1.55cm of q0] (s1) {$q_1$};
      \node[state, right=1.1cm of s1] (s2) {$q_2$};
      \node[state, right=1.1cm of s2] (s3) {$q_3$};
      \node[state, accepting, right=1.55cm of s3] (qf) {$q_F$};
      \path (q0) edge[loop above] node {$|\delta| \le 0.2$} (q0)
            (q0) edge node {$|\delta| > 0.2$} (s1)
            (s1) edge (s2)
            (s2) edge (s3)
            (s3) edge node {$|\delta| > 0.2$} (qf)
            (s1) edge[bend left=20]node[below] {$|\delta| \le 0.2$} (q0)
            (s2) edge[bend left=50]node[below] {$|\delta| \le 0.2$} (q0)
            (s3) edge[bend left=80] node[below] {$|\delta| \le 0.2$} (q0)
            (qf) edge[loop above] node {$\ast$} (qf);
    \end{tikzpicture}}
  \caption{Safety and co-safety specifications used in the experiments. Here $v$ is the
  vehicle speed (m/s) and $\delta$ the steering angle; doubly circled states are
  accepting, $q_0$ is the initial state, and $\ast$ marks a self-loop on any symbol.}  \label{fig:specs}
\end{figure}
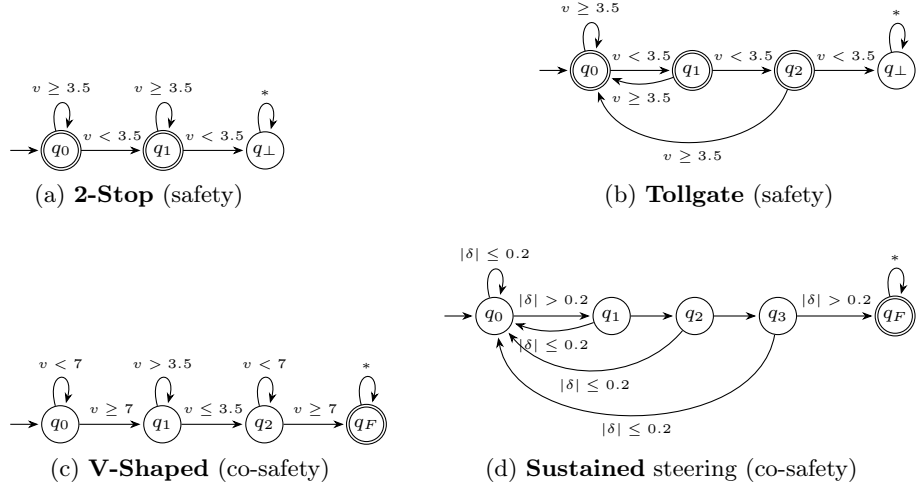

Figure \ref{fig:specs} illustrates the four specifications used in the experiments.

\newpage
\section{Full Accuracy Results}
\label{appendix:fixed_budget_results}

This section presents the full experimental results corresponding to~\Cref{subsec:exp-accuracy}.



\setlength{\tabcolsep}{10pt}
\begin{longtable}{l ccc}
\caption{Estimated satisfaction probabilities for the monolithic baseline and our compositional method under the fixed-budget regime described in \Cref{subsec:exp-accuracy}. $\Delta$ denotes the compositional estimate minus the monolithic estimate.}
\label{tab:fixed-trace-budget-full}
\\
\toprule  
Scenario & Monolithic & Compositional & $\Delta$
\\
\midrule
\endfirsthead

\multicolumn{4}{l}{\small\itshape \Cref{tab:fixed-trace-budget-full} continued} \\
\toprule  
Scenario & Monolithic & Compositional & $\Delta$
\\
\midrule
\endhead

\midrule
\multicolumn{4}{r}{\small\itshape continued on next page} \\
\endfoot

\bottomrule
\endlastfoot

\multicolumn{4}{l}{\textbf{2-Stop} (safety)} \\
\addlinespace[2pt]
\straight\seq\intersection                                          & $0.6650$ & $0.6712$ & $+0.0062$ \\
\straight\seq\intersection\seq\straight                             & $0.6000$ & $0.6700$ & $+0.0700$ \\
\straight\seq\roundabout\seq\curved                                 & $0.3890$ & $0.4511$ & $+0.0621$ \\
\curved\seq\straight\seq\intersection\seq\straight                  & $0.5380$ & $0.6156$ & $+0.0776$ \\
\curved\seq\intersection\seq\straight\seq\intersection\seq\curved   & $0.3550$ & $0.3849$ & $+0.0299$ \\
\straight \seq \cho(\curved, \intersection, \roundabout)            & $0.9770$ & $0.9511$ & $-0.0259$ \\
\straight \seq \shu(\curved, \intersection, \roundabout)            & $0.8910$ & $0.8775$ & $-0.0135$ \\
\midrule
\multicolumn{4}{l}{\textbf{Tollgate} (safety)} \\
\addlinespace[2pt]
\straight\seq\intersection                                          & $0.6750$ & $0.7074$ & $+0.0324$ \\
\straight\seq\intersection\seq\straight                             & $0.6520$ & $0.6898$ & $+0.0378$ \\
\straight\seq\roundabout\seq\curved                                 & $0.4310$ & $0.4740$ & $+0.0430$ \\
\curved\seq\straight\seq\intersection\seq\straight                  & $0.5670$ & $0.6206$ & $+0.0536$ \\
\curved\seq\intersection\seq\straight\seq\intersection\seq\curved   & $0.3820$ & $0.3819$ & $-0.0001$ \\
\straight \seq \cho(\curved, \intersection, \roundabout)            & $0.9800$ & $0.9584$ & $-0.0216$ \\
\straight \seq \shu(\curved, \intersection, \roundabout)            & $0.9150$ & $0.8941$ & $-0.0209$ \\
\midrule
\multicolumn{4}{l}{\textbf{V-Shaped} (co-safety)} \\
\addlinespace[2pt]
\straight\seq\intersection                                          & $0.3580$ & $0.3327$ & $-0.0253$ \\
\straight\seq\intersection\seq\straight                             & $0.3930$ & $0.3524$ & $-0.0406$ \\
\straight\seq\roundabout\seq\curved                                 & $0.5570$ & $0.5784$ & $+0.0214$ \\
\curved\seq\straight\seq\intersection\seq\straight                  & $0.4720$ & $0.4194$ & $-0.0526$ \\
\curved\seq\intersection\seq\straight\seq\intersection\seq\curved   & $0.6150$ & $0.6577$ & $+0.0427$ \\
\straight \seq \cho(\curved, \intersection, \roundabout)            & $0.0130$ & $0.0415$ & $+0.0285$ \\
\straight \seq \shu(\curved, \intersection, \roundabout)            & $0.3250$ & $0.3101$ & $-0.0149$ \\
\midrule
\multicolumn{4}{l}{\textbf{Sustained} (co-safety)} \\
\addlinespace[2pt]
\straight\seq\intersection                                          & $0.5680$ & $0.5305$ & $-0.0375$ \\
\straight\seq\intersection\seq\straight                             & $0.6280$ & $0.5479$ & $-0.0801$ \\
\straight\seq\roundabout\seq\curved                                 & $0.9630$ & $0.9960$ & $+0.0330$ \\
\curved\seq\straight\seq\intersection\seq\straight                  & $0.7180$ & $0.6612$ & $-0.0568$ \\
\curved\seq\intersection\seq\straight\seq\intersection\seq\curved   & $0.8420$ & $0.8729$ & $+0.0309$ \\
\straight \seq \cho(\curved, \intersection, \roundabout)            & $0.2550$ & $0.2377$ & $-0.0173$ \\
\straight \seq \shu(\curved, \intersection, \roundabout)            & $0.3100$ & $0.4262$ & $+0.1162$ \\
\end{longtable}

\newpage
\section{Full Efficiency Results}
\label{appendix:efficiency}

This section presents the complete experimental results corresponding to~\Cref{subsec:eval_eff}.
We evaluate efficiency under three settings: 
(i) both the compositional approach and the monolithic baseline are given a 15-minute wall-clock budget for trace generation (\Cref{app:time-budget-15}),
(ii) both approaches are given a 60-minute wall-clock budget for trace generation (\Cref{app:time-budget-60}), and
(iii) a compute-matched ablation in which the monolithic baseline retains the 15-minute budget, while the compositional approach receives an equal budget for each of the four primitive scenarios (\straight, \intersection, \roundabout, \curved), yielding $225$ seconds per primitive (\Cref{app:ablation-matched}).

\subsection{15-Minute Time-Budget Experiments}
\label{app:time-budget-15}

\Cref{tab:time-budget-15min} reports the compositional and monolithic estimates under a fixed 15-minute time budget for all 28 specification--scenario cells, and \Crefrange{fig:app_time_two_stop}{fig:app_time_sustained} show how these estimates converge as the budget grows.
  
\setlength{\tabcolsep}{4pt}
\begin{longtable}{l @{\hskip 16pt} r@{\hskip 10pt}r @{\hskip 16pt} r@{\hskip 10pt}r @{\hskip 16pt} r}
\caption{15-minute time-budget results. $\rho$ is the satisfaction probability estimate; $N$ is the number of traces generated; $\Delta$ denotes the compositional estimate minus the monolithic estimate.}
\label{tab:time-budget-15min}\\
\toprule
& \multicolumn{2}{c}{Monolithic} & \multicolumn{2}{c}{Compositional} & \\
\cmidrule(lr){2-3}\cmidrule(lr){4-5}
Scenario & $N$ & $\rho$ & $N$ & $\rho$ & $\Delta$ \\
\midrule
\endfirsthead

\multicolumn{6}{l}{\small\itshape \Cref{tab:time-budget-15min} continued} \\
\toprule
& \multicolumn{2}{c}{Monolithic} & \multicolumn{2}{c}{Compositional} & \\
\cmidrule(lr){2-3}\cmidrule(lr){4-5}
Scenario & $N$ & $\rho$ & $N$ & $\rho$ & $\Delta$ \\
\midrule  
\endhead

\midrule
\multicolumn{6}{r}{\small\itshape continued on next page} \\
\endfoot

\bottomrule
\endlastfoot

\multicolumn{6}{l}{\textbf{Two-stop} (safety)} \\
\addlinespace[2pt]
\straight\seq\intersection                     & $2347$ & $0.6412$ & $12262$ & $0.6558$ & $+0.0146$ \\
\straight\seq\intersection\seq\straight               & $1618$ & $0.6057$ & $12262$ & $0.6533$ & $+0.0476$ \\
\straight\seq\roundabout\seq\curved               & $788$ & $0.3820$ & $12335$ & $0.4410$ & $+0.0590$ \\
\curved\seq\straight\seq\intersection\seq\straight         & $753$ & $0.5618$ & $14871$ & $0.5892$ & $+0.0275$ \\
\curved\seq\intersection\seq\straight\seq\intersection\seq\curved   & $460$ & $0.3217$ & $14871$ & $0.3547$ & $+0.0330$ \\
\straight \seq \cho(\curved, \intersection, \roundabout)     & $49$ & $1.0000$ & $312$ & $0.9676$ & $-0.0324$ \\
\straight \seq \shu(\curved, \intersection, \roundabout)    & $20$ & $0.9500$ & $312$ & $0.9062$ & $-0.0438$ \\
\midrule
\multicolumn{6}{l}{\textbf{Tollgate} (safety)} \\
\addlinespace[2pt]
\straight\seq\intersection                     & $2347$ & $0.6860$ & $12262$ & $0.6792$ & $-0.0068$ \\
\straight\seq\intersection\seq\straight               & $1618$ & $0.6508$ & $12262$ & $0.6770$ & $+0.0262$ \\
\straight\seq\roundabout\seq\curved               & $788$ & $0.4226$ & $12335$ & $0.4638$ & $+0.0412$ \\
\curved\seq\straight\seq\intersection\seq\straight         & $753$ & $0.6069$ & $14871$ & $0.6195$ & $+0.0126$ \\
\curved\seq\intersection\seq\straight\seq\intersection\seq\curved   & $460$ & $0.3630$ & $14871$ & $0.3922$ & $+0.0292$ \\
\straight \seq \cho(\curved, \intersection, \roundabout)     & $49$ & $1.0000$ & $312$ & $0.9769$ & $-0.0231$ \\
\straight \seq \shu(\curved, \intersection, \roundabout)    & $20$ & $0.9500$ & $312$ & $0.9322$ & $-0.0178$ \\
\midrule

\multicolumn{6}{l}{\textbf{V-shaped} (co-safety, reported as $1-\rho$)} \\
\addlinespace[2pt]
\straight\seq\intersection                     & $2347$ & $0.3775$ & $12262$ & $0.3467$ & $-0.0308$ \\
\straight\seq\intersection\seq\straight               & $1618$ & $0.4017$ & $12262$ & $0.3689$ & $-0.0328$ \\
\straight\seq\roundabout\seq\curved               & $788$ & $0.5799$ & $12335$ & $0.5860$ & $+0.0061$ \\
\curved\seq\straight\seq\intersection\seq\straight         & $753$ & $0.4449$ & $14871$ & $0.4380$ & $-0.0069$ \\
\curved\seq\intersection\seq\straight\seq\intersection\seq\curved   & $460$ & $0.6783$ & $14871$ & $0.6778$ & $-0.0005$ \\
\straight \seq \cho(\curved, \intersection, \roundabout)     & $49$ & $0.0204$ & $312$ & $0.0206$ & $+0.0002$ \\
\straight \seq \shu(\curved, \intersection, \roundabout)    & $20$ & $0.6000$ & $312$ & $0.2964$ & $-0.3036$ \\
\midrule
\multicolumn{6}{l}{\textbf{Sustained} (co-safety, reported as $1-\rho$)}
\\
\addlinespace[2pt]  
\straight\seq\intersection                     & $2347$ & $0.5684$ & $12262$ & $0.4960$ & $-0.0724$ \\
\straight\seq\intersection\seq\straight               & $1618$ & $0.6193$ & $12262$ & $0.5167$ & $-0.1025$ \\
\straight\seq\roundabout\seq\curved               & $788$ & $0.9657$ & $12335$ & $0.9954$ & $+0.0296$ \\
\curved\seq\straight\seq\intersection\seq\straight         & $753$ & $0.6866$ & $14871$ & $0.6465$ & $-0.0401$ \\
\curved\seq\intersection\seq\straight\seq\intersection\seq\curved   & $460$ & $0.8565$ & $14871$ & $0.8627$ & $+0.0062$ \\
\straight \seq \cho(\curved, \intersection, \roundabout)     & $49$ & $0.2041$ & $216$ & $0.0813$ & $-0.1228$ \\
\straight \seq \shu(\curved, \intersection, \roundabout)    & $20$ & $0.5500$ & $216$ & $0.2439$ & $-0.3061$ \\
\end{longtable}

\par\medskip
\begingroup
\captionsetup{type=figure}
\captionsetup[subfloat]{justification=centering}
\centering
\subfloat[\straight\seq\intersection ($N_m{=}2347$, $N_c{=}12262$)]{%
       \includegraphics[width=0.49\linewidth,height=100pt,keepaspectratio]{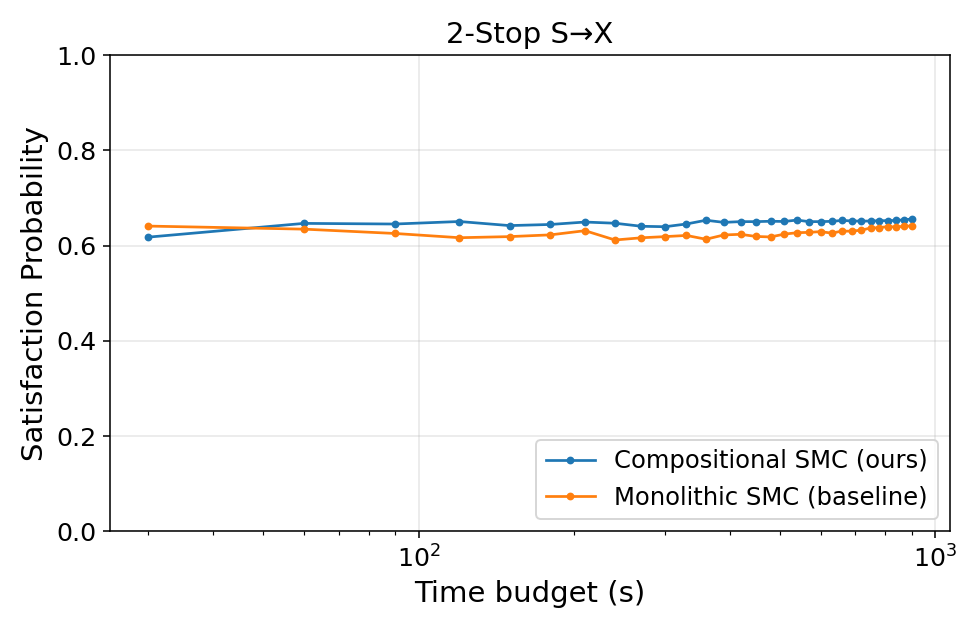}}\hfill
\subfloat[\straight\seq\intersection\seq\straight ($N_m{=}1618$, $N_c{=}12262$)]{%
       \includegraphics[width=0.49\linewidth,height=100pt,keepaspectratio]{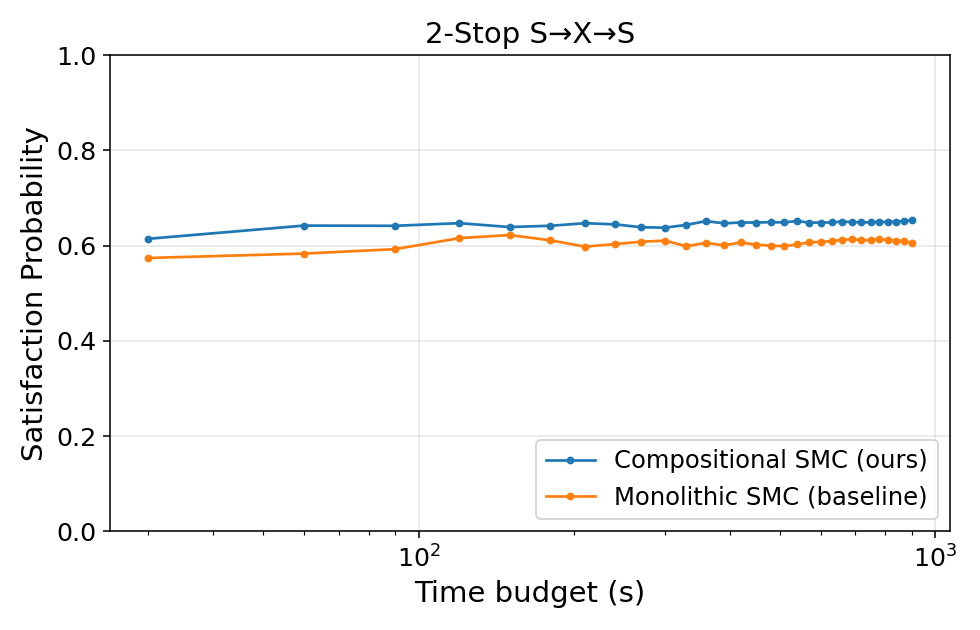}}\par\smallskip
\subfloat[\straight\seq\roundabout\seq\curved ($N_m{=}788$, $N_c{=}12335$)]{%
       \includegraphics[width=0.49\linewidth,height=100pt,keepaspectratio]{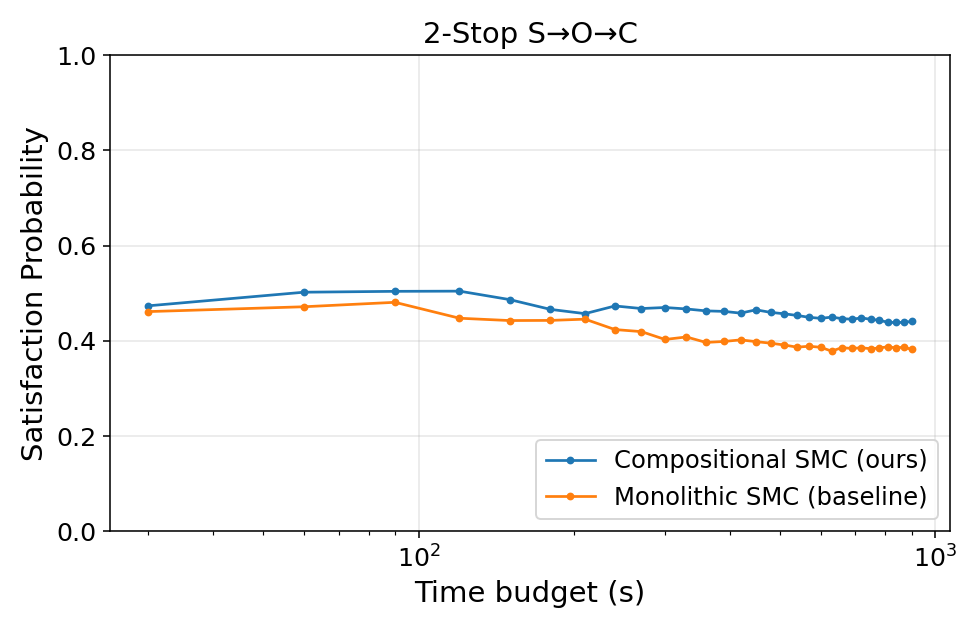}}\hfill
\subfloat[\curved\seq\straight\seq\intersection\seq\straight ($N_m{=}753$, $N_c{=}14871$)]{%
       \includegraphics[width=0.49\linewidth,height=100pt,keepaspectratio]{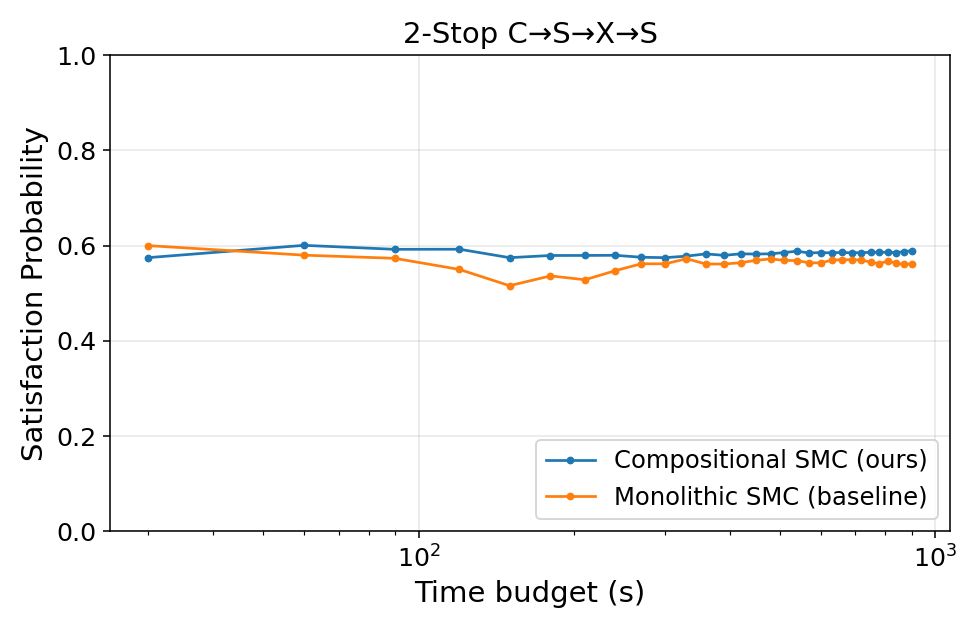}}\par\smallskip
\subfloat[\curved\seq\intersection\seq\straight\seq\intersection\seq\curved ($N_m{=}460$, $N_c{=}14871$)]{%
       \includegraphics[width=0.49\linewidth,height=100pt,keepaspectratio]{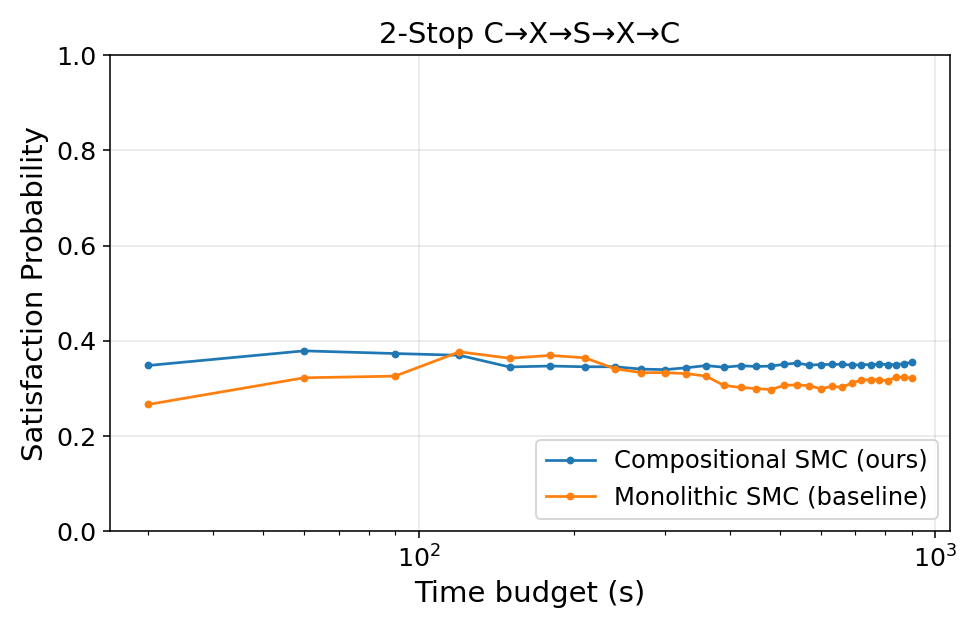}}\hfill
\subfloat[\straight \seq \cho(\curved, \intersection, \roundabout) ($N_m{=}49$, $N_c{=}312$)]{%
       \includegraphics[width=0.49\linewidth,height=100pt,keepaspectratio]{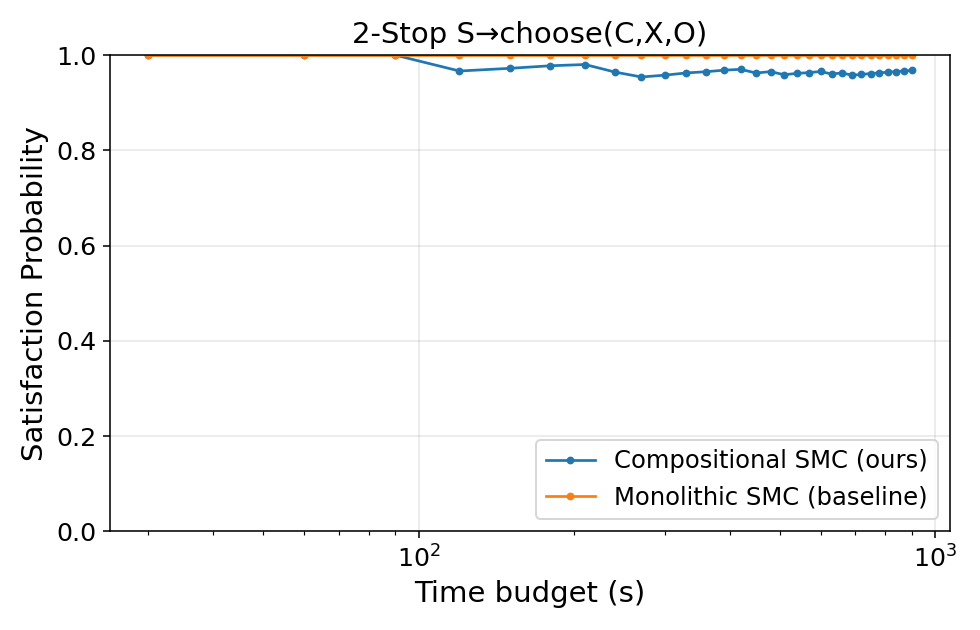}}\par\smallskip
\subfloat[\straight \seq \shu(\curved, \intersection, \roundabout) ($N_m{=}20$, $N_c{=}312$)]{%
       \includegraphics[width=0.49\linewidth,height=100pt,keepaspectratio]{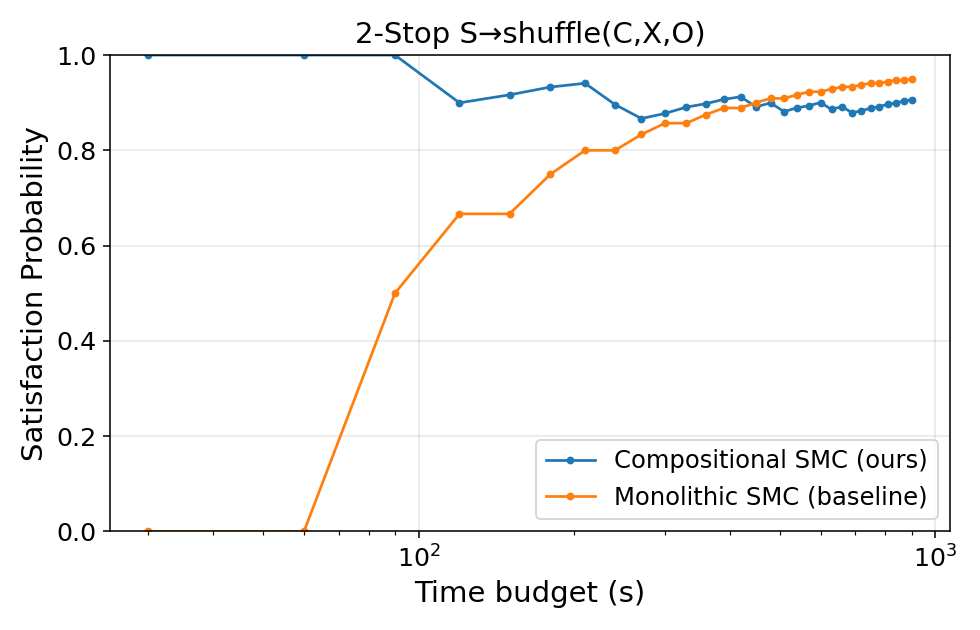}}\par\smallskip
\captionof{figure}{15-minute time-budget results for the \textbf{2-Stop} specification. $N_m$ and $N_c$ are the number of monolithic and compositional primitive traces generated within the budget.}
\label{fig:app_time_two_stop}
\endgroup
\par\medskip

\par\medskip
\begingroup
\captionsetup{type=figure}
\captionsetup[subfloat]{justification=centering}
\centering
\subfloat[\straight\seq\intersection ($N_m{=}2347$, $N_c{=}12262$)]{%
       \includegraphics[width=0.49\linewidth,height=100pt,keepaspectratio]{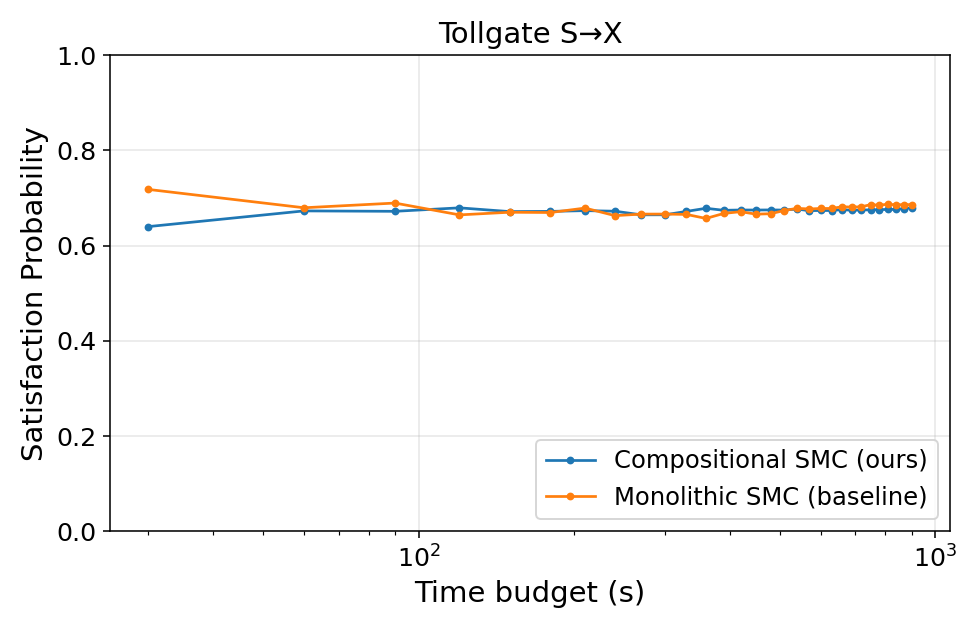}}\hfill
\subfloat[\straight\seq\intersection\seq\straight ($N_m{=}1618$, $N_c{=}12262$)]{%
       \includegraphics[width=0.49\linewidth,height=100pt,keepaspectratio]{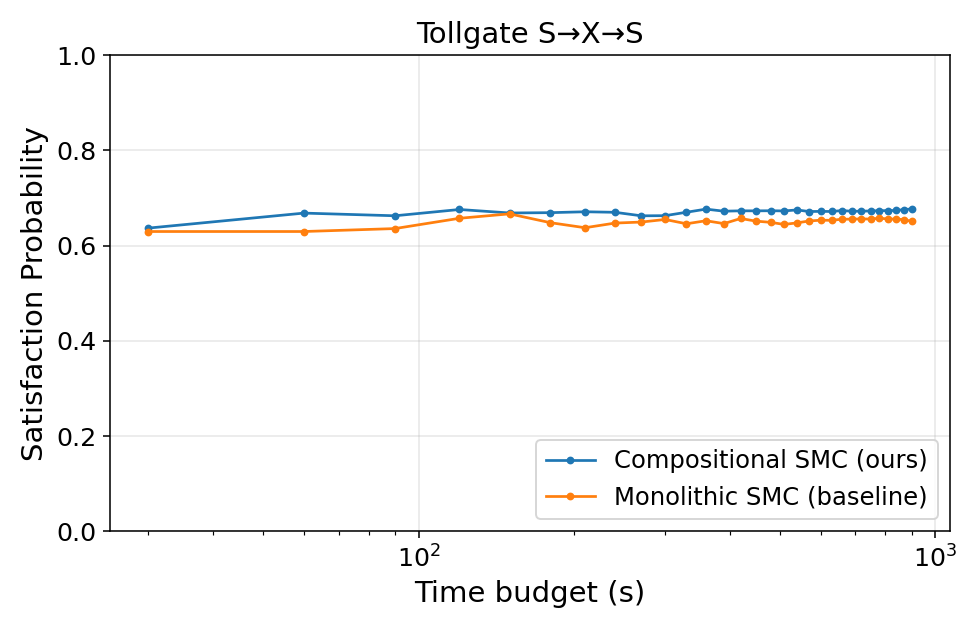}}\par\smallskip
\subfloat[\straight\seq\roundabout\seq\curved ($N_m{=}788$, $N_c{=}12335$)]{%
       \includegraphics[width=0.49\linewidth,height=100pt,keepaspectratio]{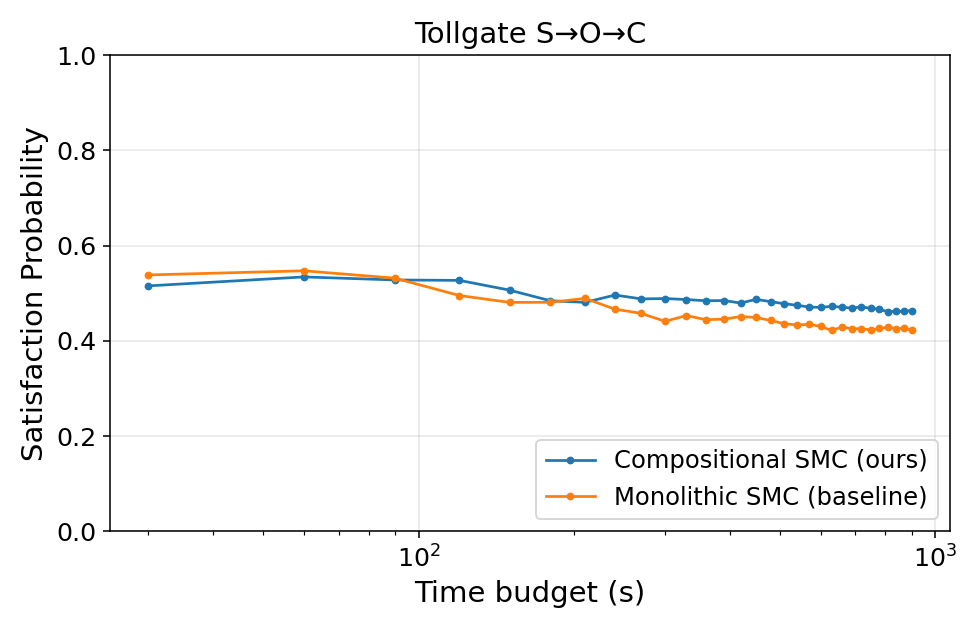}}\hfill
\subfloat[\curved\seq\straight\seq\intersection\seq\straight ($N_m{=}753$, $N_c{=}14871$)]{%
       \includegraphics[width=0.49\linewidth,height=100pt,keepaspectratio]{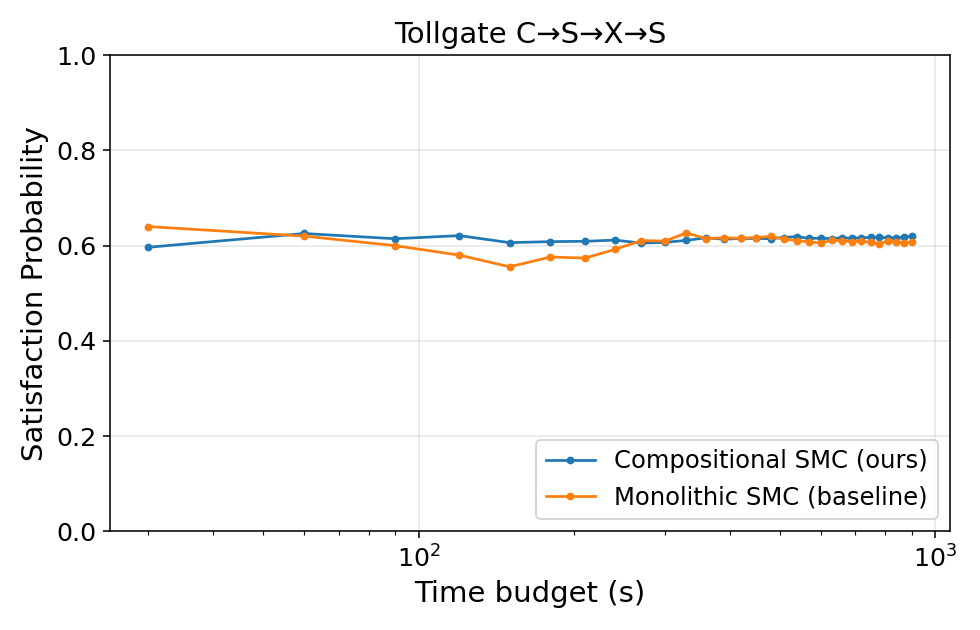}}\par\smallskip
\subfloat[\curved\seq\intersection\seq\straight\seq\intersection\seq\curved ($N_m{=}460$, $N_c{=}14871$)]{%
       \includegraphics[width=0.49\linewidth,height=100pt,keepaspectratio]{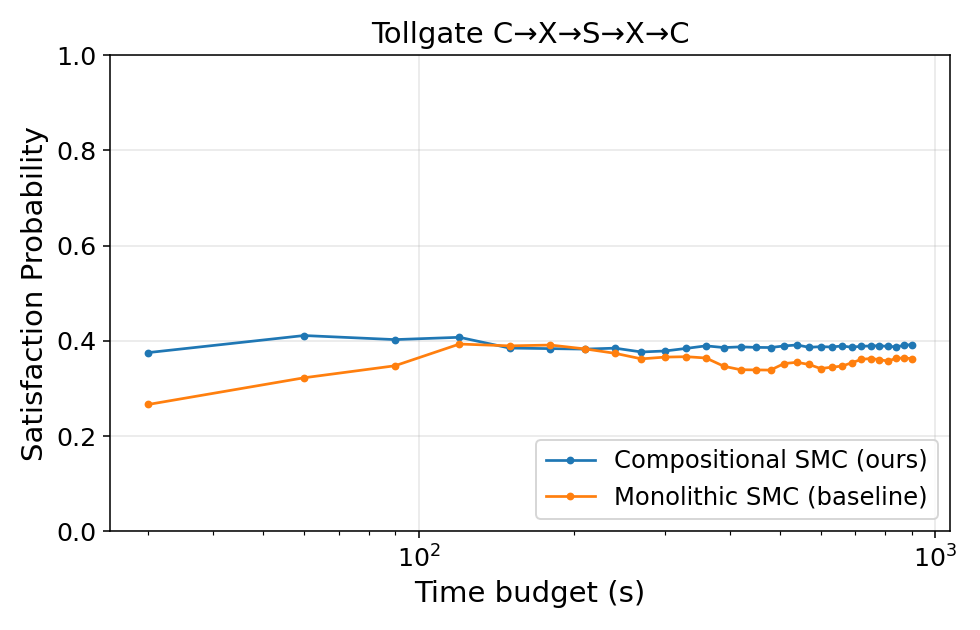}}\hfill
\subfloat[\straight \seq \cho(\curved, \intersection, \roundabout) ($N_m{=}49$, $N_c{=}312$)]{%
       \includegraphics[width=0.49\linewidth,height=100pt,keepaspectratio]{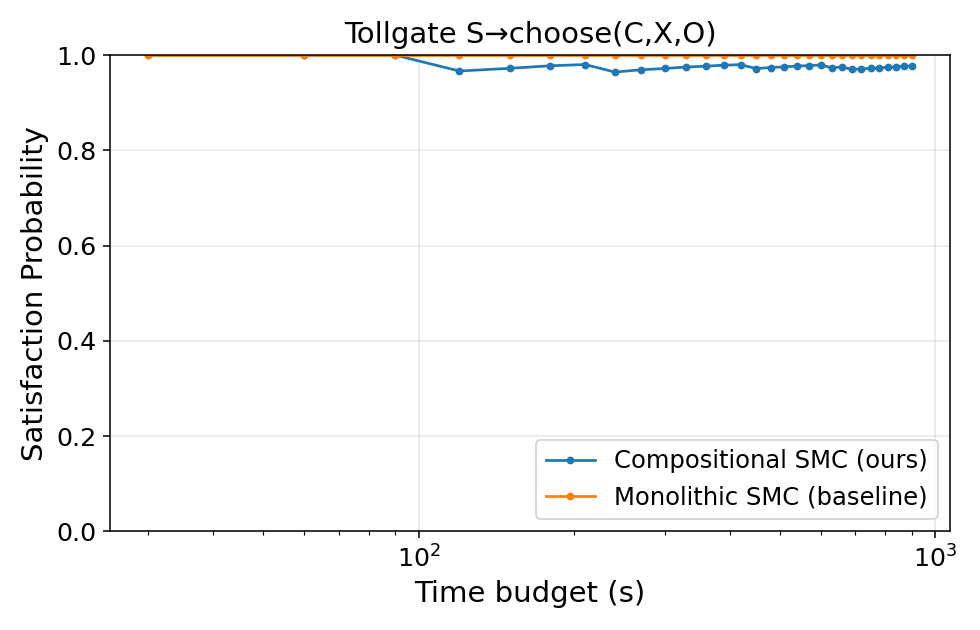}}\par\smallskip
\subfloat[\straight \seq \shu(\curved, \intersection, \roundabout) ($N_m{=}20$, $N_c{=}312$)]{%
       \includegraphics[width=0.49\linewidth,height=100pt,keepaspectratio]{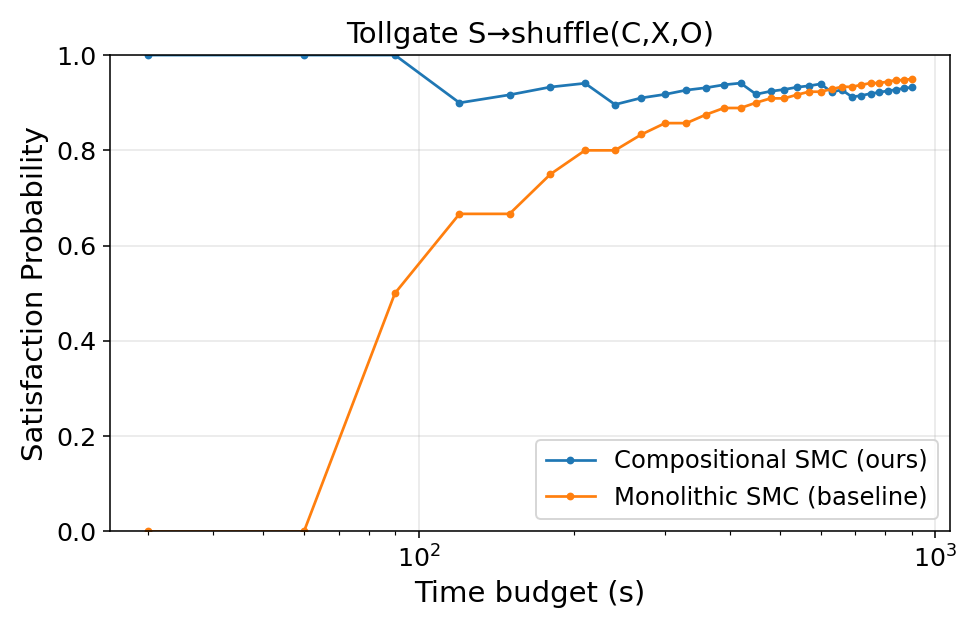}}\par\smallskip
\captionof{figure}{15-minute time-budget results for the \textbf{Tollgate} specifications. $N_m$ and $N_c$ are the number of monolithic and compositional primitive traces generated within the budget.}
\label{fig:app_time_tollgate}
\endgroup
\par\medskip

\par\medskip
\begingroup
\captionsetup{type=figure}
\captionsetup[subfloat]{justification=centering}
\centering
\subfloat[\straight\seq\intersection ($N_m{=}2347$, $N_c{=}12262$)]{%
       \includegraphics[width=0.49\linewidth,height=100pt,keepaspectratio]{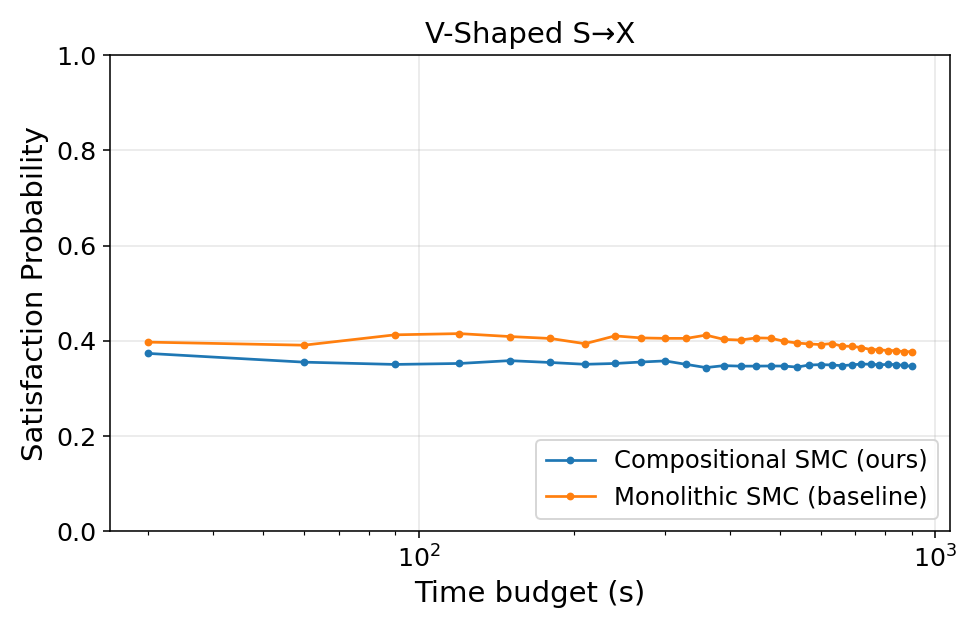}}\hfill
\subfloat[\straight\seq\intersection\seq\straight ($N_m{=}1618$, $N_c{=}12262$)]{%
       \includegraphics[width=0.49\linewidth,height=100pt,keepaspectratio]{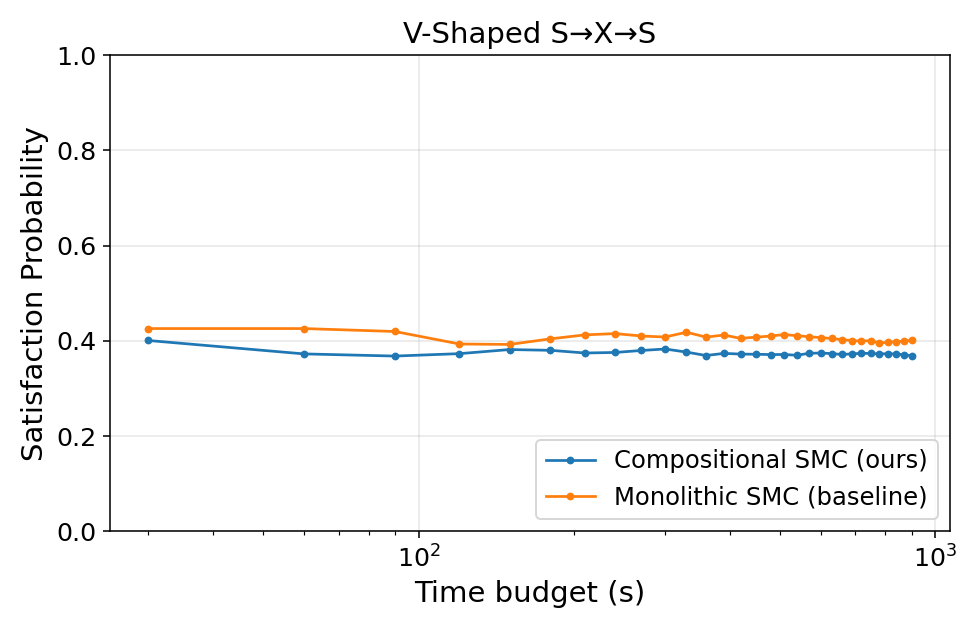}}\par\smallskip
\subfloat[\straight\seq\roundabout\seq\curved ($N_m{=}788$, $N_c{=}12335$)]{%
       \includegraphics[width=0.49\linewidth,height=100pt,keepaspectratio]{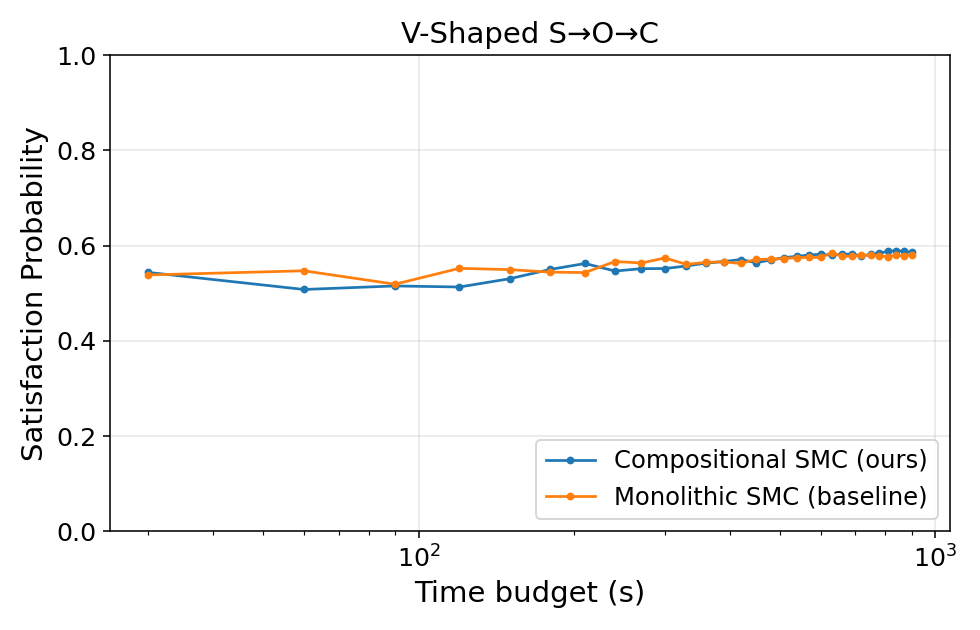}}\hfill
\subfloat[\curved\seq\straight\seq\intersection\seq\straight ($N_m{=}753$, $N_c{=}14871$)]{%
       \includegraphics[width=0.49\linewidth,height=100pt,keepaspectratio]{figures/time-budget/new/15/metadrive__vshape__CSXS.png}}\hfill
\subfloat[\curved\seq\intersection\seq\straight\seq\intersection\seq\curved ($N_m{=}460$, $N_c{=}14871$)]{%
       \includegraphics[width=0.49\linewidth,height=100pt,keepaspectratio]{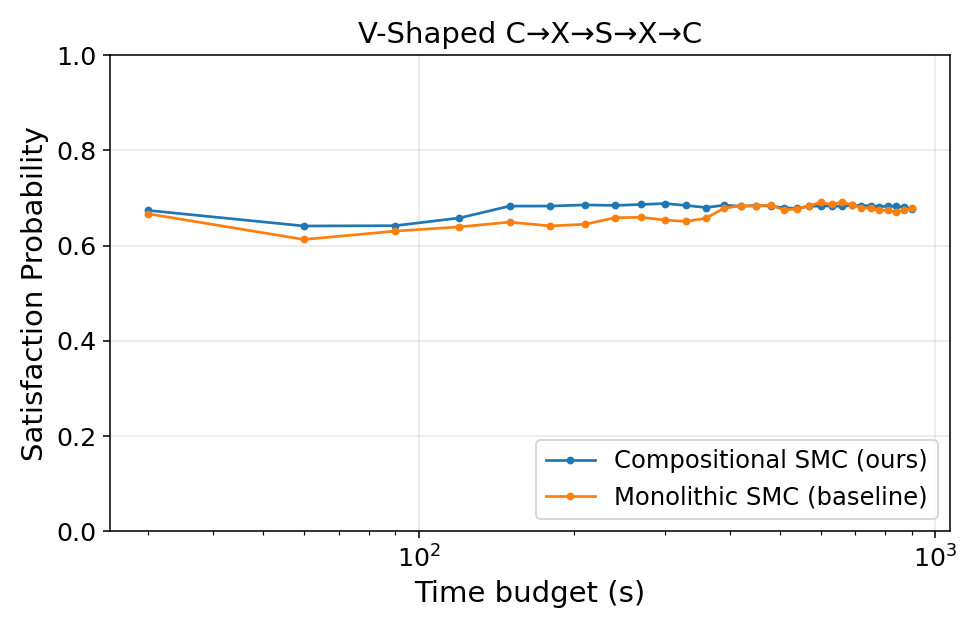}}\hfill
\subfloat[\straight \seq \cho(\curved, \intersection, \roundabout) ($N_m{=}49$, $N_c{=}312$)]{%
       \includegraphics[width=0.49\linewidth,height=100pt,keepaspectratio]{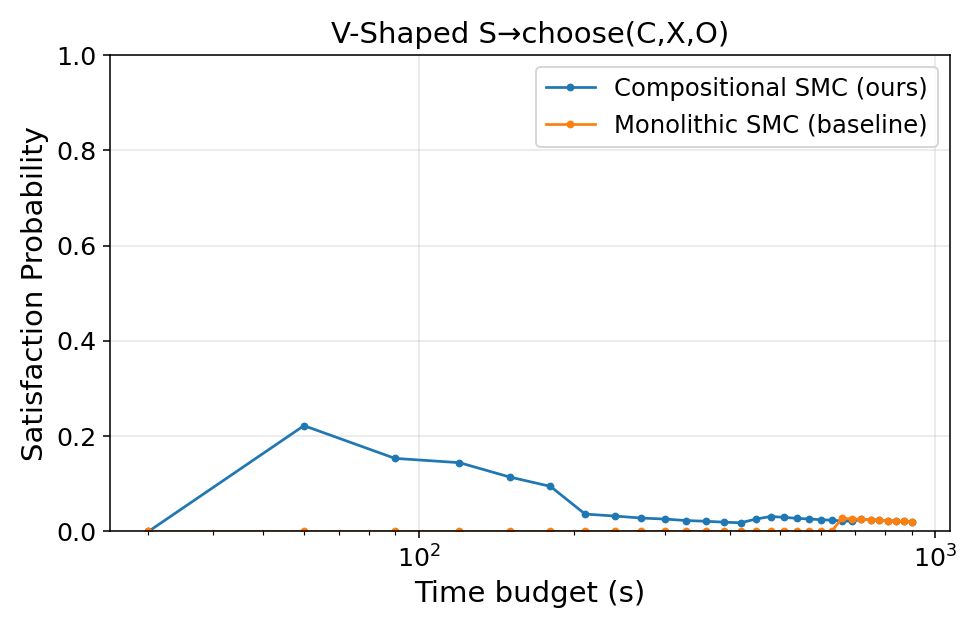}}\par\smallskip
\subfloat[\straight \seq \shu(\curved, \intersection, \roundabout) ($N_m{=}20$, $N_c{=}312$)]{%
       \includegraphics[width=0.49\linewidth,height=100pt,keepaspectratio]{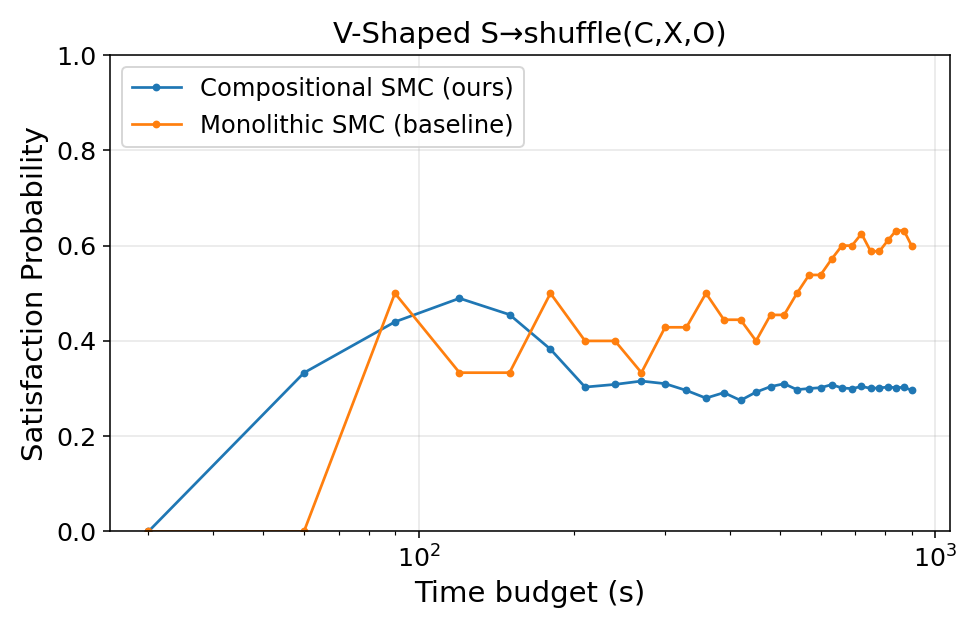}}\par\smallskip
\captionof{figure}{15-minute time-budget results for the \textbf{V-Shaped} specification. $N_m$ and $N_c$ are the number of monolithic and compositional primitive traces generated within the budget.}
\label{fig:app_time_v_shaped}
\endgroup
\par\medskip

\par\medskip
\begingroup
\captionsetup{type=figure}
\captionsetup[subfloat]{justification=centering}
\centering
\subfloat[\straight\seq\intersection ($N_m{=}2347$, $N_c{=}12262$)]{%
       \includegraphics[width=0.49\linewidth,height=100pt,keepaspectratio]{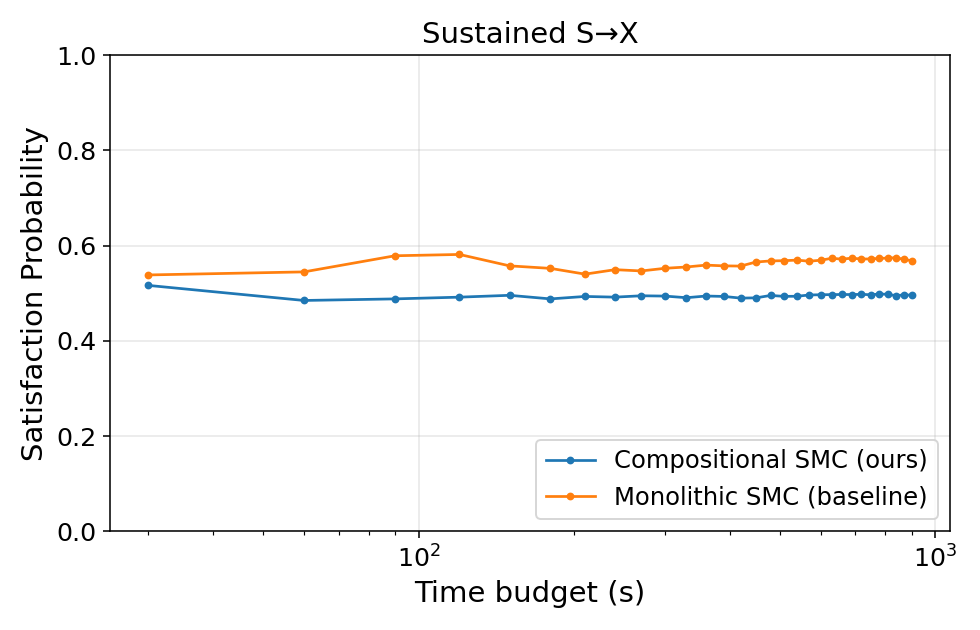}}\hfill
\subfloat[\straight\seq\intersection\seq\straight ($N_m{=}1618$, $N_c{=}12262$)]{%
       \includegraphics[width=0.49\linewidth,height=100pt,keepaspectratio]{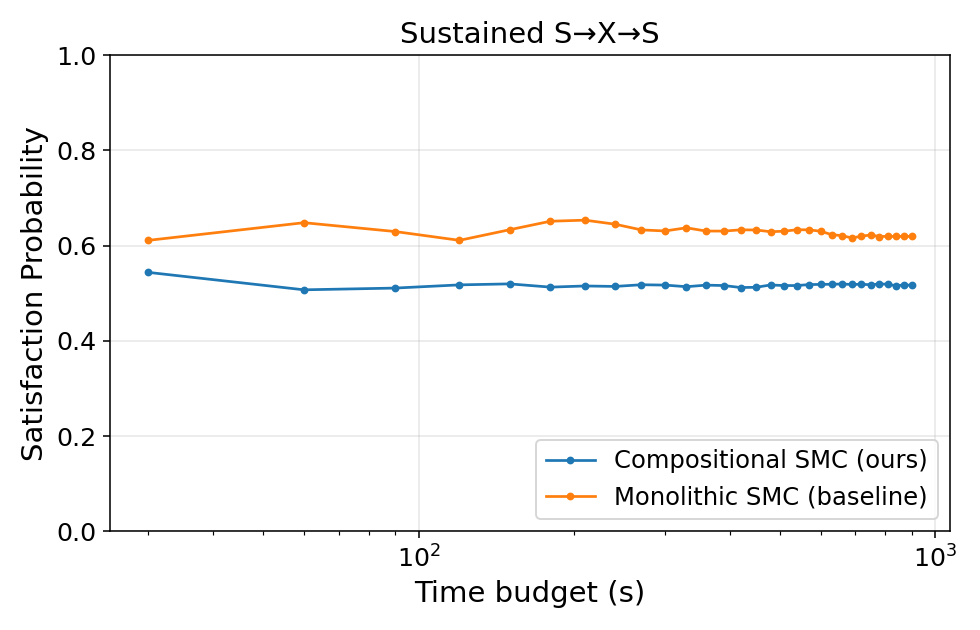}}\par\smallskip
\subfloat[\straight\seq\roundabout\seq\curved ($N_m{=}788$, $N_c{=}12335$)]{%
       \includegraphics[width=0.49\linewidth,height=100pt,keepaspectratio]{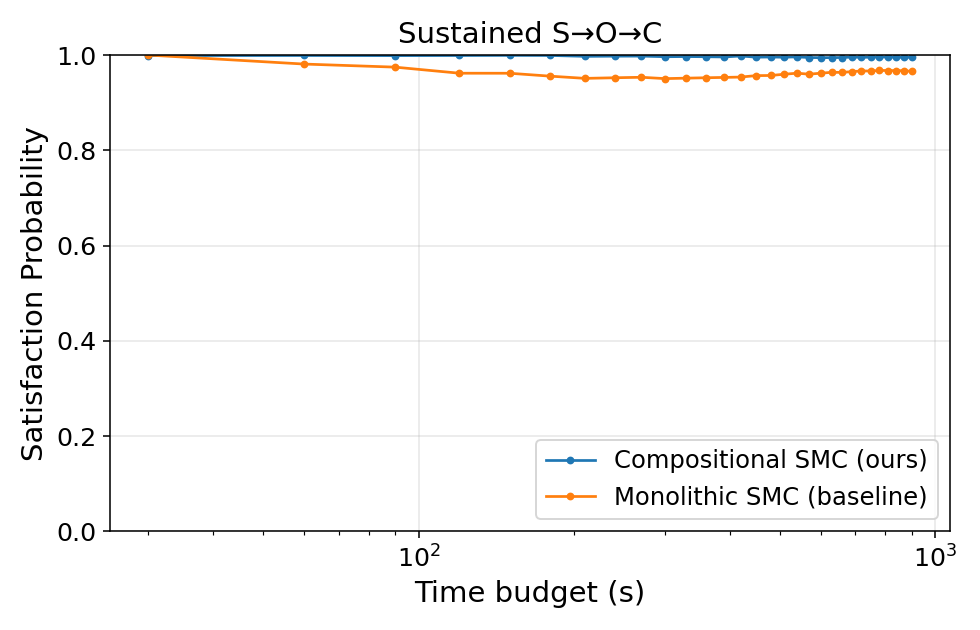}}\hfill
\subfloat[\curved\seq\straight\seq\intersection\seq\straight ($N_m{=}753$, $N_c{=}14871$)]{%
       \includegraphics[width=0.49\linewidth,height=100pt,keepaspectratio]{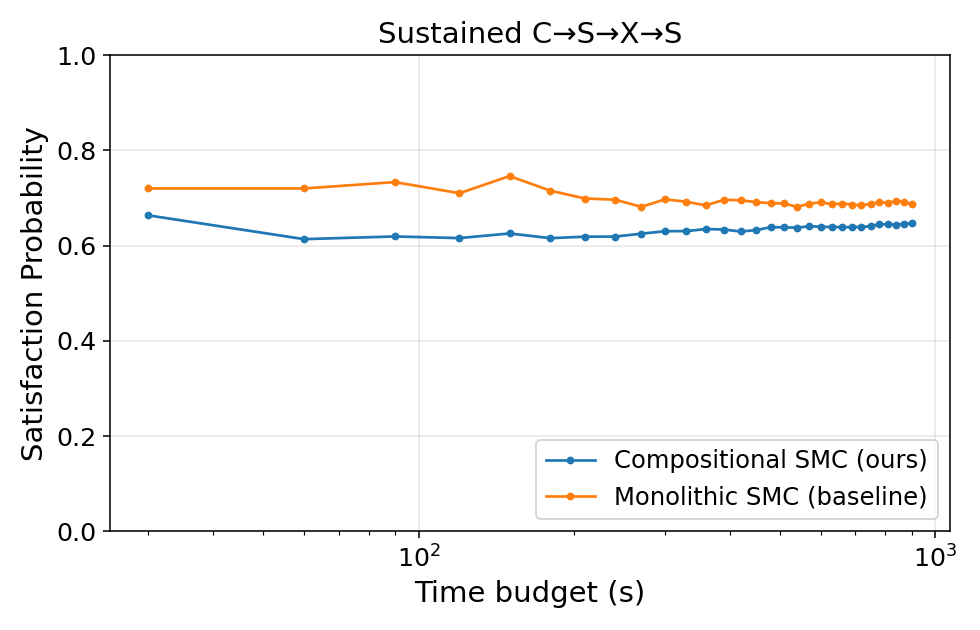}}\par\smallskip
\subfloat[\curved\seq\intersection\seq\straight\seq\intersection\seq\curved ($N_m{=}460$, $N_c{=}14871$)]{%
       \includegraphics[width=0.49\linewidth,height=100pt,keepaspectratio]{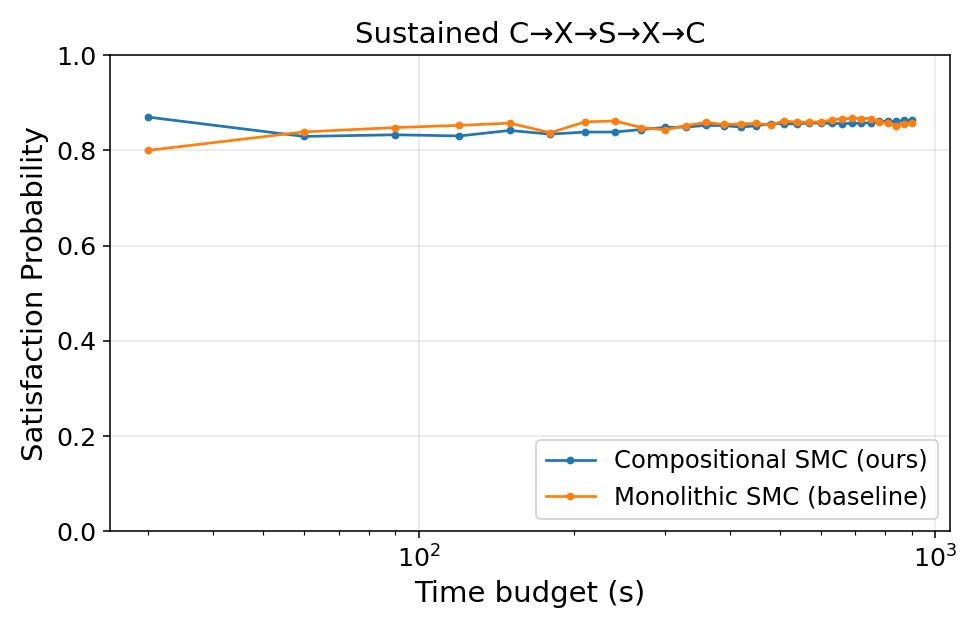}}\hfill
\subfloat[\straight \seq \cho(\curved, \intersection, \roundabout) ($N_m{=}49$, $N_c{=}216$)]{%
       \includegraphics[width=0.49\linewidth,height=100pt,keepaspectratio]{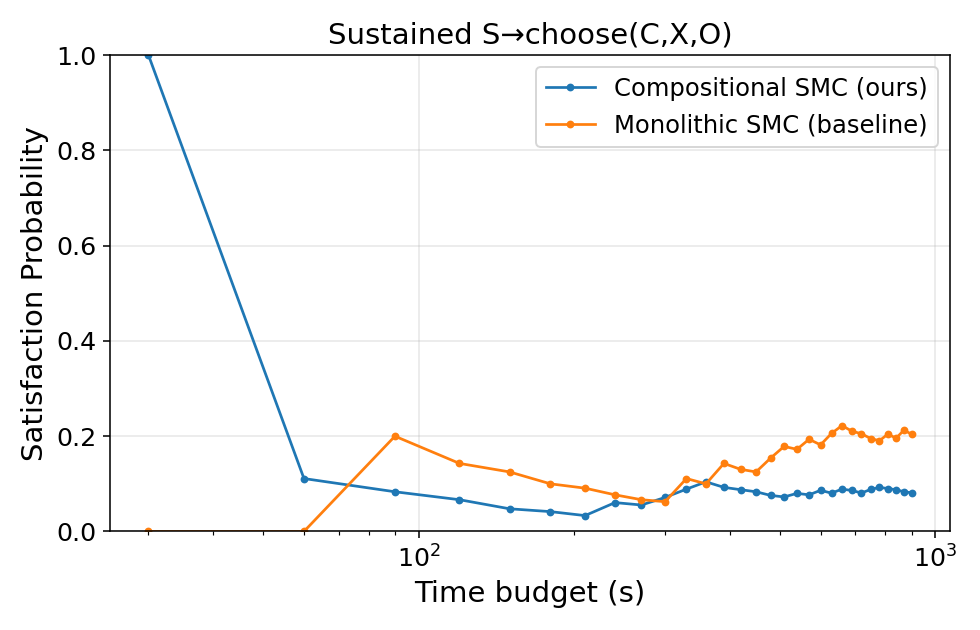}}\par\smallskip
\subfloat[\straight \seq \shu(\curved, \intersection, \roundabout) ($N_m{=}20$, $N_c{=}216$)]{%
       \includegraphics[width=0.49\linewidth,height=100pt,keepaspectratio]{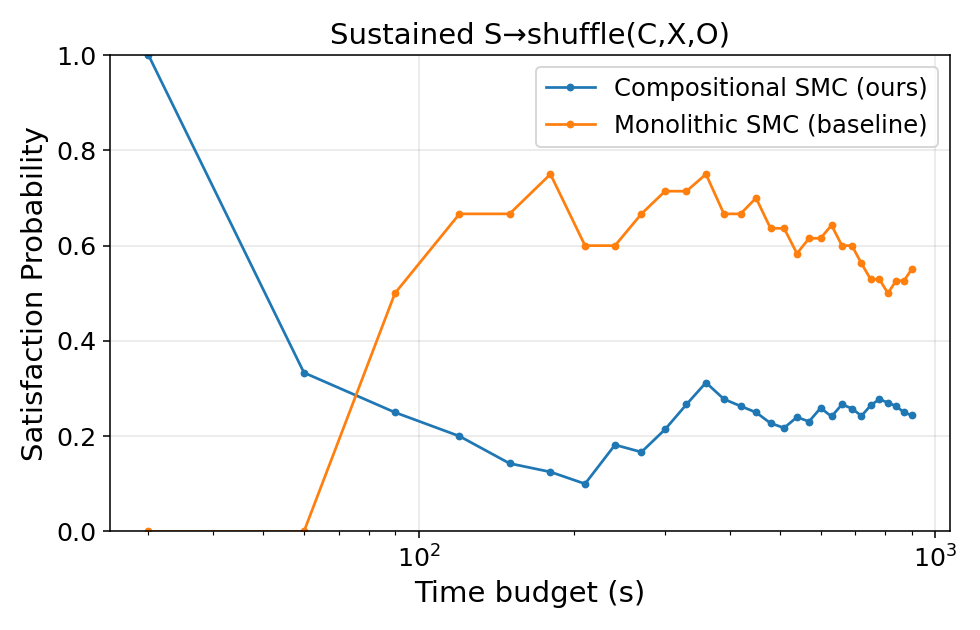}}\par\smallskip
\captionof{figure}{15-minute time-budget results for the \textbf{Sustained} specification. $N_m$ and $N_c$ are the number of monolithic and compositional primitive traces generated within the budget.}
\label{fig:app_time_sustained}
\endgroup
\par\medskip

As described in \Cref{subsec:eval_eff}, the monolithic estimate deviates from each method's own
fixed-trace-budget reference (\Cref{tab:fixed-trace-budget-full}) by $0.039$ against $0.029$ for the
compositional one. The gap is concentrated among scenarios involving \cho and \shu ($0.089$ vs.\ $0.059$), where the $15$-minute monolithic budget buys as few as $20$ traces.


\subsection{60-Minute Time-Budget Experiments}
\label{app:time-budget-60}

To obtain a sharper monolithic baseline for the sample-starved \cho/\shu scenarios, we increase the time budget to 60 minutes. \Cref{tab:time-budget-60min} reports the resulting estimates and \Cref{fig:app_time60} shows their convergence.
  
\setlength{\tabcolsep}{4pt}
\begin{longtable}{l @{\hskip 16pt} r@{\hskip 10pt}r @{\hskip 16pt} r@{\hskip 10pt}r @{\hskip 16pt} r}
\caption{60-minute time-budget results on the \cho/\shu scenarios. $\rho$ is the satisfaction probability estimate; $N$ is the number of traces generated; $\Delta$ denotes the compositional estimate minus
the monolithic estimate.}
\label{tab:time-budget-60min}\\
\toprule
& \multicolumn{2}{c}{Monolithic} & \multicolumn{2}{c}{Compositional} & \\
\cmidrule(lr){2-3}\cmidrule(lr){4-5}
Scenario & $N$ & $\rho$ & $N$ & $\rho$ & $\Delta$ \\
\midrule
\endfirsthead

\multicolumn{6}{l}{\small\itshape \Cref{tab:time-budget-60min} continued} \\
\toprule
& \multicolumn{2}{c}{Monolithic} & \multicolumn{2}{c}{Compositional} & \\
\cmidrule(lr){2-3}\cmidrule(lr){4-5}
Scenario & $N$ & $\rho$ & $N$ & $\rho$ & $\Delta$ \\
\midrule
\endhead

\midrule
\multicolumn{6}{r}{\small\itshape continued on next page} \\
\endfoot

\bottomrule
\endlastfoot

\multicolumn{6}{l}{\textbf{Two-stop} (safety)} \\
\addlinespace[2pt]  
\straight \seq \cho(\curved, \intersection, \roundabout)     & $188$ & $0.9734$ & $1161$ & $0.9503$ & $-0.0231$ \\
\straight \seq \shu(\curved, \intersection, \roundabout)    & $76$ & $0.9211$ & $1161$ & $0.8795$ & $-0.0415$ \\
\midrule
\multicolumn{6}{l}{\textbf{Tollgate} (safety)} \\
\addlinespace[2pt]
\straight \seq \cho(\curved, \intersection, \roundabout)     & $188$ & $0.9787$ & $1161$ & $0.9568$ & $-0.0219$ \\
\straight \seq \shu(\curved, \intersection, \roundabout)    & $76$ & $0.9211$ & $1161$ & $0.8964$ & $-0.0246$ \\
\midrule
\multicolumn{6}{l}{\textbf{V-shaped} (co-safety, reported as $1-\rho$)} \\
\addlinespace[2pt]
\straight \seq \cho(\curved, \intersection, \roundabout)     & $188$ & $0.0532$ & $1161$ & $0.0509$ & $-0.0023$ \\
\straight \seq \shu(\curved, \intersection, \roundabout)    & $76$ & $0.5132$ & $1161$ & $0.3490$ & $-0.1642$ \\
\midrule
\multicolumn{6}{l}{\textbf{Sustained steering} (co-safety, reported as $1-\rho$)}
\\
\addlinespace[2pt]  
\straight \seq \cho(\curved, \intersection, \roundabout)     & $188$ & $0.2234$ & $816$ & $0.1229$ & $-0.1005$ \\
\straight \seq \shu(\curved, \intersection, \roundabout)    & $76$ & $0.5132$ & $816$ & $0.3665$ & $-0.1467$ \\
\end{longtable}

\par\medskip
\begingroup
\captionsetup{type=figure}
\captionsetup[subfloat]{justification=centering}
\centering
\subfloat[\textbf{2-Stop} on \straight \seq \cho(\curved, \intersection, \roundabout) ($N_m{=}188$, $N_c{=}1161$)]{%
       \includegraphics[width=0.49\linewidth,height=100pt,keepaspectratio]{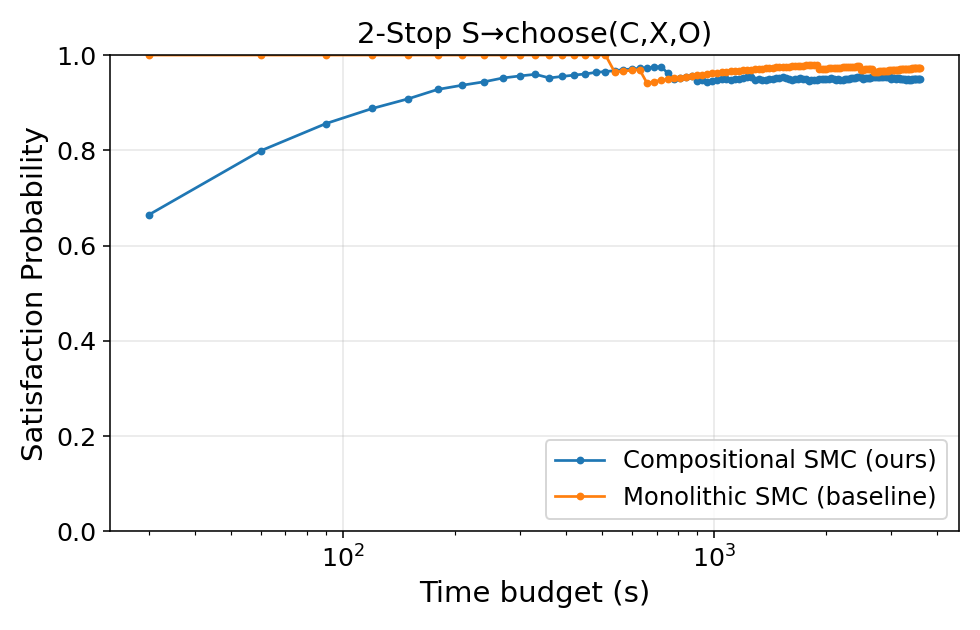}}\hfill
\subfloat[\textbf{2-Stop} on \straight \seq \shu(\curved, \intersection, \roundabout) ($N_m{=}76$, $N_c{=}1161$)]{%
       \includegraphics[width=0.49\linewidth,height=100pt,keepaspectratio]{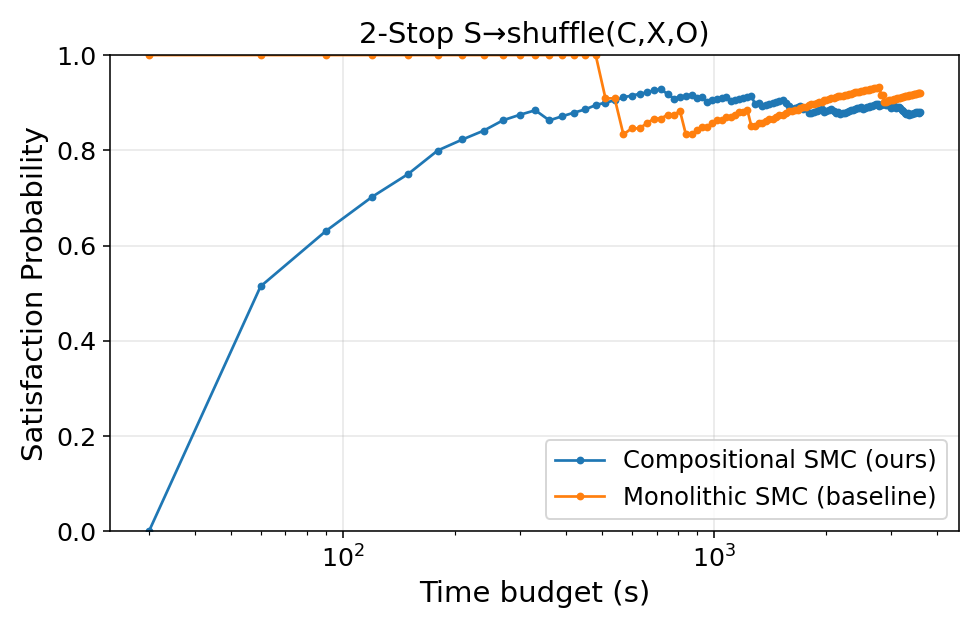}}\par\smallskip
\subfloat[\textbf{Tollgate} on \straight \seq \cho(\curved, \intersection, \roundabout) ($N_m{=}188$, $N_c{=}1161$)]{%
       \includegraphics[width=0.49\linewidth,height=100pt,keepaspectratio]{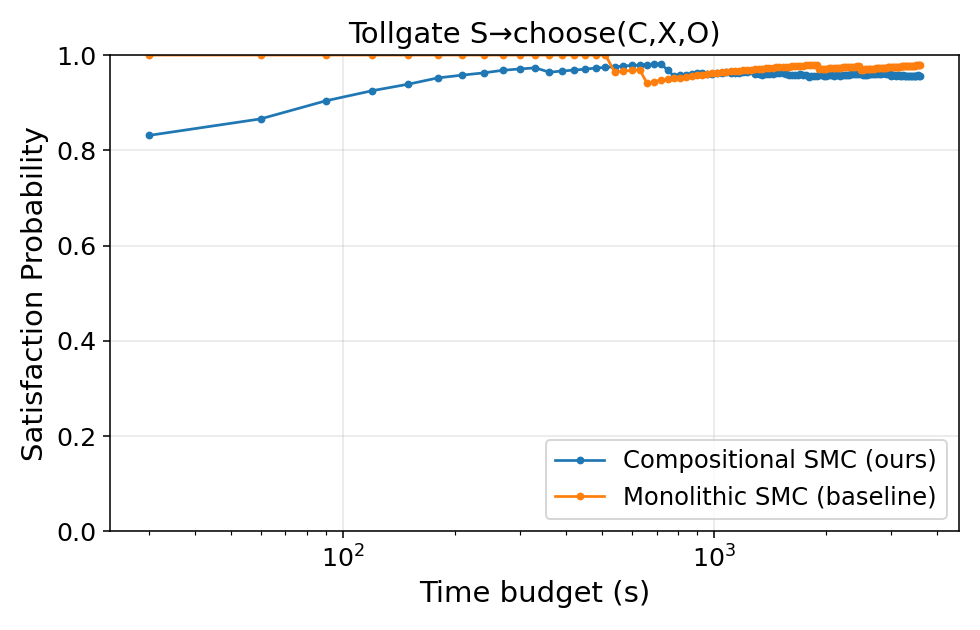}}\hfill
\subfloat[\textbf{Tollgate} on \straight \seq \shu(\curved, \intersection, \roundabout) ($N_m{=}76$, $N_c{=}1161$)]{%
       \includegraphics[width=0.49\linewidth,height=100pt,keepaspectratio]{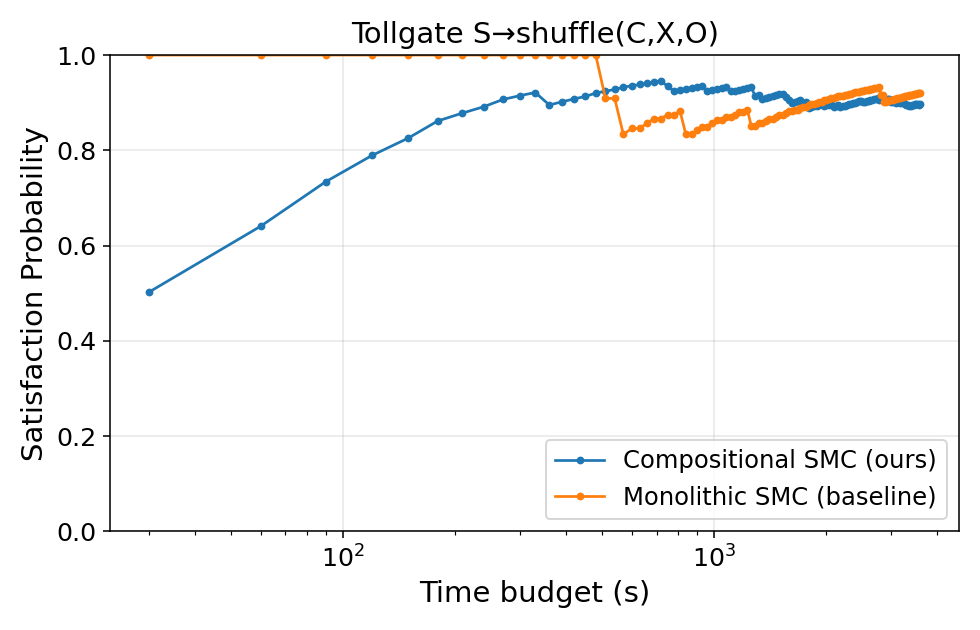}}\par\smallskip
\subfloat[\textbf{V-shaped} on \straight \seq \cho(\curved, \intersection, \roundabout) ($N_m{=}188$, $N_c{=}1161$)]{%
       \includegraphics[width=0.49\linewidth,height=100pt,keepaspectratio]{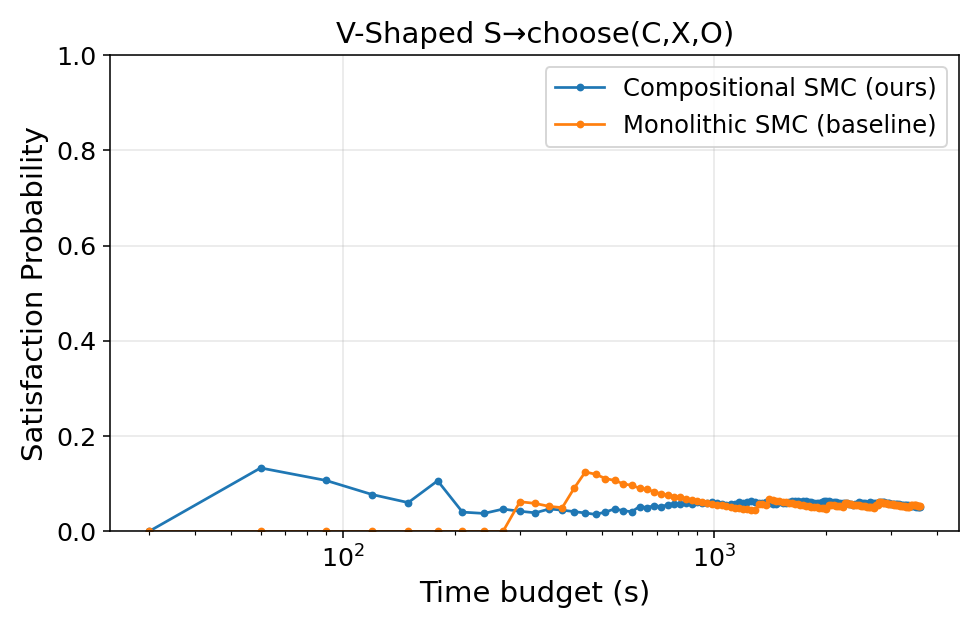}}\hfill
\subfloat[\textbf{V-shaped} on \straight \seq \shu(\curved, \intersection, \roundabout) ($N_m{=}76$, $N_c{=}1161$)]{%
       \includegraphics[width=0.49\linewidth,height=100pt,keepaspectratio]{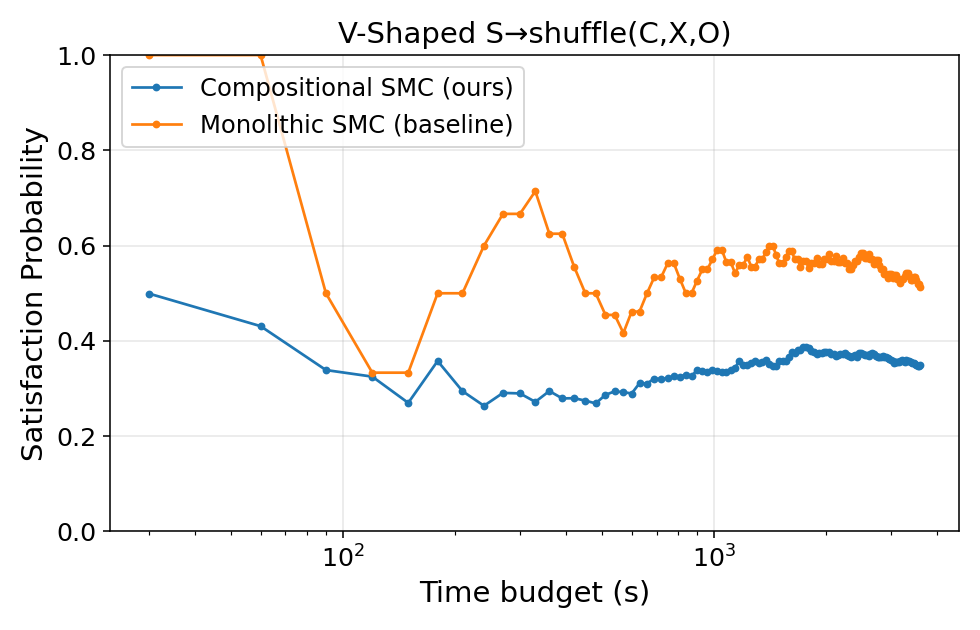}}\par\smallskip
\subfloat[\textbf{Sustained} on \straight \seq \cho(\curved, \intersection, \roundabout) ($N_m{=}188$, $N_c{=}816$)]{%
       \includegraphics[width=0.49\linewidth,height=100pt,keepaspectratio]{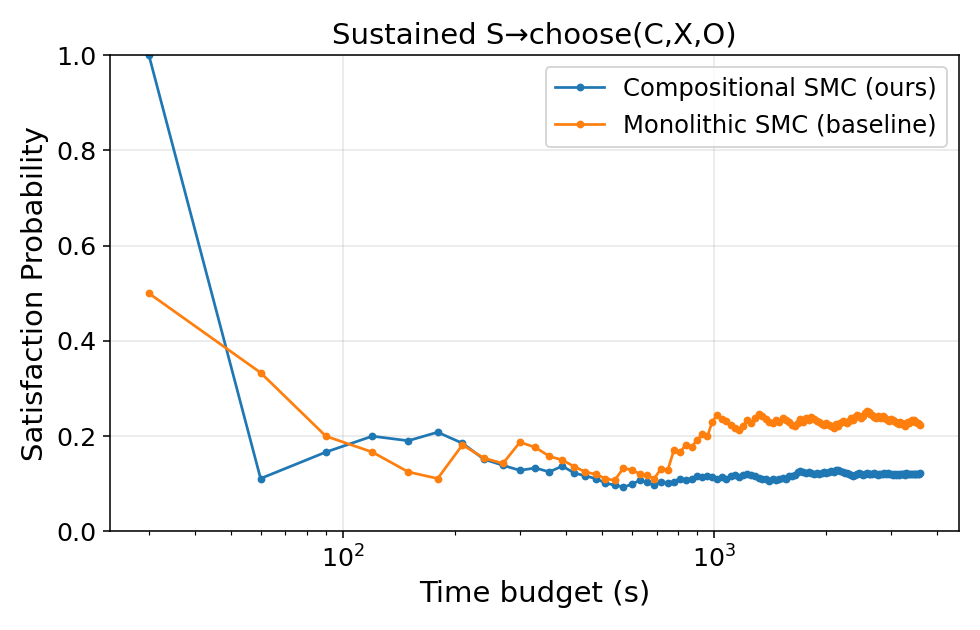}}\hfill
\subfloat[\textbf{Sustained} on \straight \seq \shu(\curved, \intersection, \roundabout) ($N_m{=}76$, $N_c{=}816$)]{%
       \includegraphics[width=0.49\linewidth,height=100pt,keepaspectratio]{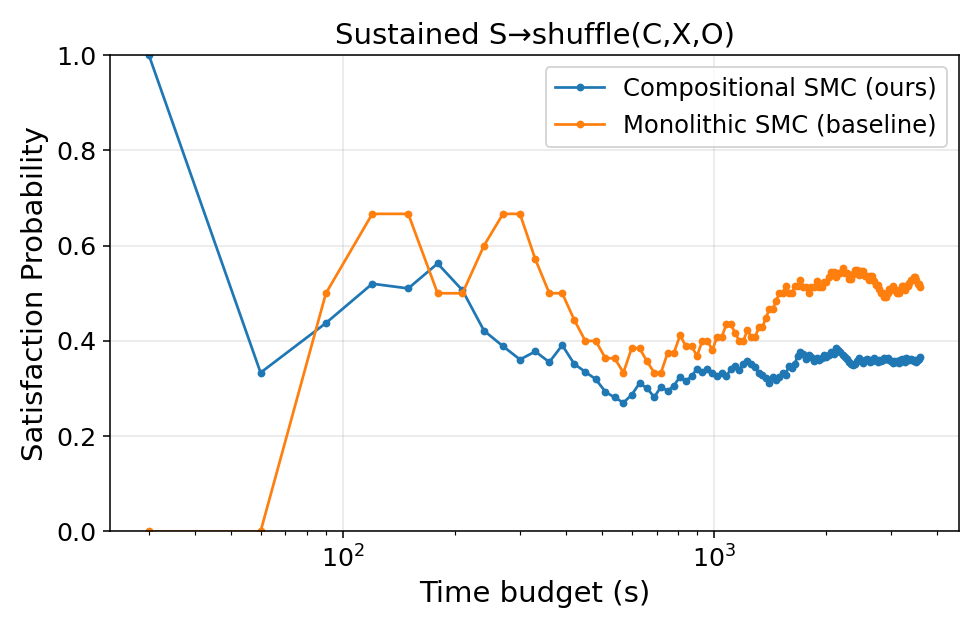}}\par\smallskip
\captionof{figure}{60-minute time-budget results on the \cho/\shu scenarios. Figures (a)--(d) are the two safety specifications (\textbf{2-Stop}, \textbf{Tollgate}). Figures (e)--(h) are the two co-safety specifications (\textbf{V-shaped}, \textbf{Sustained}), reported as $1-\rho$. $N_m$ and $N_c$ are the number of monolithic and compositional primitive traces generated within the budget.}
\label{fig:app_time60}
\endgroup
\par\medskip

With four times the budget, the monolithic estimate becomes more accurate but still does not catch up. Averaged across the eight scenario--specification pairs, its deviation from the reference decreases from $0.089$ at $15$ minutes to $0.063$, whereas the compositional deviation drops from $0.059$ to $0.029$, leaving the monolithic estimate roughly twice as far from the reference. The remaining gap is concentrated in the two co-safety \shu pairs: even at $N_m=76$, the monolithic estimate remains $0.188$ (\textbf{V-Shaped}) and $0.203$ (\textbf{Sustained}) from its reference, compared with $0.039$ and $0.060$, respectively, for the compositional method. For \textbf{V-Shaped} \shu, whose two reference values differ by only $0.015$, this provides a clear convergence comparison: the compositional estimate is already within $0.039$ of the shared reference value, while the $60$-minute monolithic run remains $0.188$ away.


\subsection{Compute-Matched Ablation}
\label{app:ablation-matched}


A natural concern is whether the compositional method's advantage is genuine or merely
a consequence of sample count: within a fixed wall-clock window it generates many short
primitive traces, whereas the monolithic baseline generates only a few long ones. To
control for this, we level the comparison by giving both methods the same total
simulation budget rather than the same wall-clock time, charging the compositional
method only the cost of generating a single monolithic scenario and asking whether it
still recovers the same estimates.

Each primitive pool $P_i$ ($i=1,\dots,k$) contains $N_i$ traces generated within budget $T$, equal to the budget for one monolithic scenario. Splitting the same budget evenly across the $k$ primitives allocates $T/k$ to each. Assuming a roughly constant per-trace cost, this corresponds to the first $\lfloor N_i/k \rfloor$ traces in each pool. We therefore truncate every pool to this prefix, preserving a total budget of $T$ while leaving the estimator unchanged.

\setlength{\tabcolsep}{4pt}
\begin{longtable}{l @{\hskip 16pt} r@{\hskip 10pt}r @{\hskip 16pt} r@{\hskip 10pt}r @{\hskip 16pt} r}
\caption{Compute-matched ablation. $\rho$ is the satisfaction probability estimate; $N$ is the number of traces generated; $\Delta$ denotes the compositional estimate minus the monolithic estimate.}
\label{tab:ablation_matched}\\
\toprule
& \multicolumn{2}{c}{Monolithic} & \multicolumn{2}{c}{\shortstack{Compositional\\(matched)}} & \\
\cmidrule(lr){2-3}\cmidrule(lr){4-5}
Scenario & $N$ & $\rho$ & $N$ & $\rho$ & $\Delta$ \\
\midrule
\endfirsthead

\multicolumn{6}{l}{\small\itshape \Cref{tab:ablation_matched} continued} \\
\toprule
 & \multicolumn{2}{c}{\shortstack{Monolithic SMC\\(baseline)}} & \multicolumn{2}{c}{\shortstack{Compositional SMC\\(ours, matched)}} & \\
\cmidrule(lr){2-3}\cmidrule(lr){4-5}
Scenario & $N$ & $\rho$ & $N$ & $\rho$ & $\Delta$ \\
\midrule
\endhead

\midrule
\multicolumn{6}{r}{\small\itshape continued on next page} \\
\endfoot

\bottomrule
\endlastfoot

\multicolumn{6}{l}{\textbf{Two-stop} (safety)} \\
\addlinespace[2pt]
\straight\seq\intersection                     & $2347$ & $0.6412$ & $3064$ & $0.6482$ & $+0.0070$ \\
\straight\seq\intersection\seq\straight               & $1618$ & $0.6057$ & $3064$ & $0.6459$ & $+0.0402$ \\
\straight\seq\roundabout\seq\curved               & $788$ & $0.3820$ & $3082$ & $0.4617$ & $+0.0797$ \\
\curved\seq\straight\seq\intersection\seq\straight         & $753$ & $0.5618$ & $3716$ & $0.5781$ & $+0.0163$ \\
\curved\seq\intersection\seq\straight\seq\intersection\seq\curved   & $460$ & $0.3217$ & $3716$ & $0.3422$ & $+0.0205$ \\
\straight \seq \cho(\curved, \intersection, \roundabout)     & $49$ & $1.0000$ & $77$ & $0.9623$ & $-0.0377$ \\
\straight \seq \shu(\curved, \intersection, \roundabout)    & $20$ & $0.9500$ & $77$ & $0.8905$ & $-0.0595$ \\
\midrule
\multicolumn{6}{l}{\textbf{Tollgate} (safety)} \\
\addlinespace[2pt]
\straight\seq\intersection                     & $2347$ & $0.6860$ & $3064$ & $0.6736$ & $-0.0124$ \\
\straight\seq\intersection\seq\straight               & $1618$ & $0.6508$ & $3064$ & $0.6714$ & $+0.0206$ \\
\straight\seq\roundabout\seq\curved               & $788$ & $0.4226$ & $3082$ & $0.4837$ & $+0.0611$ \\
\curved\seq\straight\seq\intersection\seq\straight         & $753$ & $0.6069$ & $3716$ & $0.6086$ & $+0.0017$ \\
\curved\seq\intersection\seq\straight\seq\intersection\seq\curved   & $460$ & $0.3630$ & $3716$ & $0.3798$ & $+0.0168$ \\
\straight \seq \cho(\curved, \intersection, \roundabout)     & $49$ & $1.0000$ & $77$ & $0.9623$ & $-0.0377$ \\
\straight \seq \shu(\curved, \intersection, \roundabout)    & $20$ & $0.9500$ & $77$ & $0.8905$ & $-0.0595$ \\
\midrule
\multicolumn{6}{l}{\textbf{V-shaped} (co-safety, reported as $1-\rho$)} \\
\addlinespace[2pt]
\straight\seq\intersection                     & $2347$ & $0.3775$ & $3064$ & $0.3510$ & $-0.0265$ \\
\straight\seq\intersection\seq\straight               & $1618$ & $0.4017$ & $3064$ & $0.3755$ & $-0.0262$ \\
\straight\seq\roundabout\seq\curved               & $788$ & $0.5799$ & $3082$ & $0.5580$ & $-0.0219$ \\
\curved\seq\straight\seq\intersection\seq\straight         & $753$ & $0.4449$ & $3716$ & $0.4483$ & $+0.0034$ \\
\curved\seq\intersection\seq\straight\seq\intersection\seq\curved   & $460$ & $0.6783$ & $3716$ & $0.6895$ & $+0.0112$ \\
\straight \seq \cho(\curved, \intersection, \roundabout)     & $49$ & $0.0204$ & $77$ & $0.0346$ & $+0.0142$ \\
\straight \seq \shu(\curved, \intersection, \roundabout)    & $20$ & $0.6000$ & $77$ & $0.3045$ & $-0.2955$ \\
\midrule
\multicolumn{6}{l}{\textbf{Sustained} (co-safety, reported as $1-\rho$)} \\
\addlinespace[2pt]
\straight\seq\intersection                     & $2347$ & $0.5684$ & $3064$ & $0.4922$ & $-0.0762$ \\
\straight\seq\intersection\seq\straight               & $1618$ & $0.6193$ & $3064$ & $0.5140$ & $-0.1053$ \\
\straight\seq\roundabout\seq\curved               & $788$ & $0.9657$ & $3082$ & $0.9975$ & $+0.0318$ \\
\curved\seq\straight\seq\intersection\seq\straight         & $753$ & $0.6866$ & $3716$ & $0.6191$ & $-0.0675$ \\
\curved\seq\intersection\seq\straight\seq\intersection\seq\curved   & $460$ & $0.8565$ & $3716$ & $0.8389$ & $-0.0176$ \\
\straight \seq \cho(\curved, \intersection, \roundabout)     & $49$ & $0.2041$ & $53$ & $0.0333$ & $-0.1708$ \\
\straight \seq \shu(\curved, \intersection, \roundabout)    & $20$ & $0.5500$ & $53$ & $0.1000$ & $-0.4500$ \\
\end{longtable}


With the budget split evenly across the $k$ primitive pools (here, $k=4$, yielding $225$ seconds per pool), the compositional advantage in raw sample count disappears, yet its accuracy remains essentially unchanged. Relative to their respective $1{,}000$-trace references (\Cref{tab:fixed-trace-budget-full}), the compositional and monolithic estimates achieve nearly identical mean deviations across the $28$ pairs: $0.038$ and $0.039$, respectively. Thus, the compositional method retains its accuracy without relying on additional traces, while a single truncated primitive pool can still be reused across all composite scenarios, unlike the monolithic baseline, which must regenerate traces from scratch for each scenario.


\end{document}